\documentclass[11pt,a4paper]{revtex4}

\pdfoutput=1

\usepackage{amsmath}
\usepackage{color}
\usepackage{graphicx}
\usepackage{url}
\usepackage{epsfig}
\usepackage[T1]{fontenc}
\usepackage{multirow}

\usepackage{grffile}
\usepackage{amsmath,amsthm,amssymb,bm,amsfonts}
\usepackage{slashed}
\usepackage[utf8]{inputenc}

\usepackage{color}
\usepackage[normalem]{ulem}

\def\qbar{{\overline{q}}}

\def\aprime{a^\prime}

\def\kperp{{k_T}}
\def\kdwaperp{k^2_T}

\def\kperp{k_T}
\def\pperp{p_T}

\newcommand{\be}{\begin{eqnarray}}
\newcommand{\ee}{\end{eqnarray}}

\newcommand{\bea}{\begin{align}}
\newcommand{\eea}{\end{align}}

\def\kperp{{k_T}}

\def\sp{S}
\def\s0{\sigma_0}
\def\alphas{\alpha_{\mathrm{s}}}

\def\tHi{d\sigma ^{(qg^* \to q\gamma)}}
\def\qval{q_{\mathrm{val}}}

\def\be{\begin{equation}}
\def\ee{\end{equation}}
\renewcommand{\vec}[1]{\mbox{\boldmath $#1$}}

\begin{document}

\title{Prompt photon production in proton collisions as a probe of parton scattering in high energy limit}

\author{Krzysztof Golec-Biernat$^{1}$\footnote{golec@ifj.edu.pl}, Leszek Motyka$^{2}$\footnote{leszek.motyka@uj.edu.pl} and Tomasz Stebel$^{2}$\footnote{tomasz.stebel@uj.edu.pl}}
\affiliation{$^1$Institute of Nuclear Physics PAN, Radzikowskiego 152, 31-342 Krak\'ow, Poland\\
$^2$Institute of Theoretical Physics, Jagiellonian University, S.\L{}ojasiewicza 11, 30-348 Krak\'ow, Poland}

\begin{abstract}
We study the prompt photon hadroproduction at the LHC with the $k_T$-factorization approach and the $qg^* \to q\gamma$ and $g^*g^* \to q\bar q\gamma$ partonic channels, using three unintegrated gluon distributions which depend on gluon transverse momentum. They represent three different theoretical schemes which are usually considered in the $k_T$-factorization approach, known under the acronyms: KMR, CCFM and GBW gluon distributions. 
We find sensitivity of the calculated prompt photon transverse momentum distribution to the gluon transverse momentum distribution. The predictions obtained with the three approaches are compared to data, that allows to differentiate between them. We also discuss the significance of the two partonic channels, confronted with the expectations which are based on the applicability of the $k_T$-factorization scheme in the high energy approximation.
\end{abstract}

\maketitle

\section{Introduction}

Prompt photon production in hadron collisions is one of the cleanest probes of the strong interactions and the structure of hadrons. Prompt photons are produced in hard partonic scattering and  with sufficiently large photon transverse momentum $q_{T}$ and suitable photon isolation criteria, this process may be calculated with high precision in perturbative QCD framework. Since the produced photons do not experience the final state interactions, their cross sections carry information about the properties of incoming partons.

The analysis performed in this study is based on the $k_T$-factorization framework \cite{Gribov:1984tu,Catani:1990xk,Catani:1990eg} which is well motivated in the high energy limit. For the prompt photon production at the LHC in the central rapidity region, the high energy limit provides a very good approximation at a lower range of the measured photon transverse momenta. In order to apply the high energy factorization scheme it is necessary that the high energy approximation works. It has been argued that the conditions required to apply the high energy approximation hold in the quasi-multi Regge kinematics, extended for hard processes  towards the region of moderate parton momentum fraction $x$ \cite{Deak:2008ky,Motyka:2016lta}. In the small~$x$ regime, the gluon channel contributions to hadronic cross sections are dominant, and due to an intense gluon radiation, a sizable parton transverse momentum is expected to build up in the QCD radiation process. Within the $k_T$-factorization approach, such effects may be treated within all-order resummation schemes like  e.g.\ the Balitsky--Fadin--Kuraev--Lipatov (BFKL)~\cite{Fadin:1975cb,Kuraev:1977fs,Balitsky:1978ic} or Catani--Ciafaloni--Fiorani--Marche\-sini (CCFM) \cite{Ciafaloni:1987ur, Catani:1989sg,Catani:1989yc,Marchesini:1994wr} schemes. The possibility to investigate the effects of both the  all order resummations and  large transverse momenta of partons at  the tree level parton scattering made this approach vivid and fruitful for phenomenological applications.

The main purpose of this study is to investigate in detail the constraints that the prompt photon production at the Large Hadron Collider (LHC) impose on the transverse momentum dependent gluon distribution $F_g(x,\vec{k}_T,\mu_F)$, computed in the $k_T$-factorization framework at small and moderate values of $x$, in  wide ranges of the gluon momentum $k_T$ and  hard factorization scales $\mu_F$. This is an important issue since in inclusive cross sections, like e.g.\ for the Deep Inelastic Scattering (DIS), the gluon transverse momentum  is integrated out whereas  the photon transverse momentum distribution, $d\sigma^{\gamma} / dq_{T}$, is expected to exhibit significant sensitivity to the transverse momenta of incoming partons. Also the range of scales $\mu_F$ probed in the prompt photon production at the LHC is wider than in most processes used for this purpose in the past. Hence, we aim to use the new available precision data from the LHC to  better constrain the transverse momentum dependence of the gluon distribution $F_g$.

In general terms, a similar approach has already been used in the past \cite{Kopeliovich:1995an,Kopeliovich:1998nw, JalilianMarian:2012bd,Benic:2016yqt,Benic:2016uku,Benic:2018hvb,Ducloue:2017kkq,
Baranov:2007np,Baranov:2008sr,Lipatov:2011eg,Lipatov:2016wgr,Kniehl:2011hc,Santos:2020nqy,Goncalves:2020tvh}. 
In the approach proposed in \cite{Kopeliovich:1995an,Kopeliovich:1998nw} one considers a fast quark propagating through a color field of a target that scatters by gluon absorption, and a Brehmsstrahlung photon is emitted. In the high energy limit, this process may be represented in a color dipole form \cite{Kopeliovich:1995an,Kopeliovich:1998nw}, similarly to the forward Drell--Yan production \cite{Brodsky:1996nj}. This approach has an advantage to include multiple scattering effects within the framework of the Color Glass Condensate (CGC) \cite{McLerran:1993ni,McLerran:1993ka,McLerran:1994vd}, as it was done in Refs.\ \cite{
JalilianMarian:2012bd,Benic:2016yqt,Benic:2016uku,Benic:2018hvb,Ducloue:2017kkq,Santos:2020nqy,Goncalves:2020tvh}. 
Another approach, valid in the single scattering (leading twist) approximation, including effects of the partons' transverse momenta, was developed in Refs.\ \cite{Baranov:2007np,Baranov:2008sr,Lipatov:2011eg,Lipatov:2016wgr,Kniehl:2011hc}. Calculations made in both the approaches may be divided according to the final state produced in the hard scatter: the prompt photon may be produced in association with one or two jets, and possibly without an accompanying jet, as the incoming partons carry non-zero transverse momentum. The last contribution is, however, suppressed for large transverse momentum of the photon and it is often neglected. Recently, in the framework of CGC, the process of $g^*g^* \to q\bar q \gamma$ was thoroughly studied \cite{Benic:2016yqt,Benic:2016uku,Benic:2018hvb} as the leading one in hadronic and nuclei collisions at large energies, where the large gluon densities at small $x$ contribute.

One of the main goals of this paper is to study in detail the $g^*g^*$ channel in the single scattering approximation, that is very accurate at larger values of $q_T$, and to compare the results with recent data from the LHC. In addition, this analysis is an important step in our ongoing program to constrain the transverse momentum parton distributions with the LHC data, see Ref.\ \cite{Motyka:2016lta} for the analysis of the Lam--Tung relation breaking in $Z^0$ hadroproduction at the LHC. We improve on the previously known results by carefully comparing two possible realizations of the prompt photon production at the parton level with off-shell gluons $g^*$:  the $2\to 2$ processes $qg^* \to q\gamma$ and $\qbar g^* \to \qbar\gamma$, and the  $2\to 3$ process $g^*g^* \to q\bar q\gamma$. We impose in our analysis the photon isolation criteria of Frixione \cite{Frixione:1998jh} and use the data from the LHC \cite{Aad:2010sp,Chatrchyan:2011ue,Aad:2016xcr}.

Hence, besides constraining  the gluon distribution $F_g(x,\vec{k}_T,\mu_F)$, we also focus on the comparison of two different partonic channels that should be close to each other at the leading logarithmic approximation for small and moderate values of parton momentum fractions $x$. These are already mentioned, the $qg^* \to q\gamma$ and $g^*g^* \to q\bar q\gamma$ channels with off-shell gluons. In general, the gluon distribution is larger than the sea quark distributions, and due to larger anomalous dimensions, the gluon distribution is more rapid and drives the evolution of the sea quark sea distributions in scale $\mu_F$ and in parton~$x$. Hence, one may approximate the sea quarks as coming from the gluon in the last splitting of the parton evolution. This approximation is a basis of a very successful dipole picture of high energy scattering, see e.g.\ \cite{GolecBiernat:1998js,Golec-Biernat:2017lfv}. Thus, one expects that the contribution with sea quarks, $q_{\mathrm{sea}}g^* \to q\gamma$, should be well represented as a part of the $g^*g^* \to q\bar q\gamma$ contribution. 
This should hold true because the amplitude for $g^*g^* \to q\bar q\gamma$ contains diagrams describing the $g^* \to q_{\mathrm{sea}}$ splitting in the last step, and in addition subleading terms in the collinear limit. So, in principle one could expect an improved theoretical precision of the hard matrix element in the latter approach. Jumping to the conclusions, to some surprise, we find the opposite to be true. We view this result as an interesting theoretical puzzle and a strong inspiration to perform the complete NLO analysis of the process in the $k_T$-factorization framework.

The prompt photon production at the LHC has been thoroughly analyzed within the collinear factorization framework up to the NNLO accuracy \cite{Chen:2019zmr}. Using this approach,  good agreement with the data was found. We do not expect  better agreement in our analysis with the $k_T$-factorization approach since  the process has been treated so far only at the tree level. We rather apply more phenomenologically minded logic  to use the theoretically clean observables and precision data in order to refine details of the transverse momentum distribution of partons within the $k_T$-factorization framework. Nevertheless, even at the tree level, we find a rather good description of the LHC data in one of the considered scenarios. 

 The paper is organized as follows. In section \ref{Sec:2} we present the formalism for $qg^* \to q\gamma$ and $g^*g^* \to q\bar q\gamma$ channels. In section \ref{Sec:3} we discuss the transverse momentum dependent gluon distributions used in the paper. In section \ref{Sec:4} we discuss the  photon isolation criteria and present our results. We conclude in section \ref{Sec:5}.

\section{Overview of the process description}
\label{Sec:2}

\begin{figure}
\centering
\includegraphics[width=0.35\textwidth]{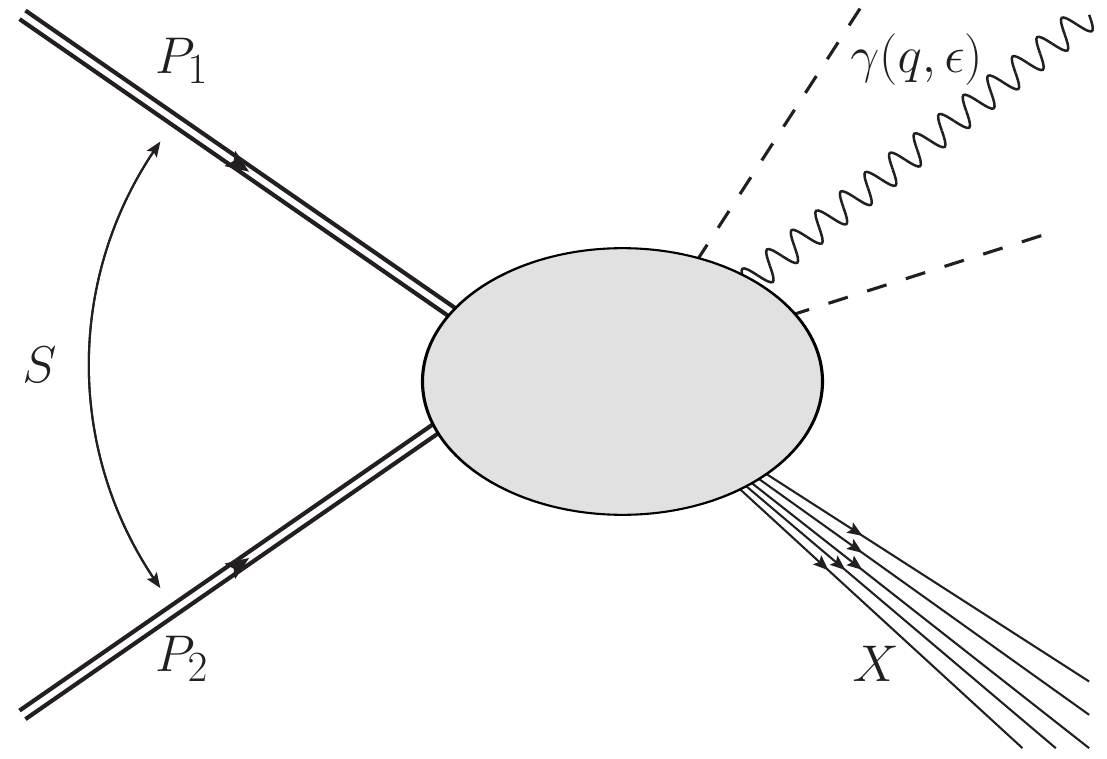}
\caption{Prompt photon production in proton-proton collisions. The dashed lines represent the photon isolation cone.}
\label{fig:1}
\end{figure}

We consider the prompt photon production in the $pp$ scattering in  the $k_T$-factorization approach,  depicted in Fig.\ \ref{fig:1}. The incoming proton beam momenta, $P_1$ and $P_2$, are consider  to be light-like: $P_1^2 = P_2^2=0$ in the high energy approximation: $S = (P_1 + P_2)^2\gg 4m_p^2$. The photon kinematics is parametrized with the help of Feynman variable $x_F$ and the transverse momentum $\vec{q}_{T}$. In the light-cone variables, the real photon momentum in the $pp$ center-of-mass frame reads
\begin{equation}
q_{\gamma} = (q^+_{\gamma},q^-_{\gamma}, \vec{q}_{T}) = (x_F \sqrt{S}, \vec{q}^2_{T}/x_F\sqrt{S},\vec{q}_{T})\, ,
\end{equation}
where $q_{\gamma}^{\pm} = q_{\gamma}^0 \pm q_{\gamma}^3$. 

At the leading order (LO), the real photon may be produced in the $2\to 2$ partonic channels: $q g \to q\gamma$, $\qbar g \to \qbar\gamma$ and $q\bar q \to g\gamma$. At the NLO,  these $2\to 2$ processes receive one loop correction and the tree level $2\to 3$ processes appear: $gg \to q\bar q \gamma$, $qg \to qg\gamma$,  $\bar qg \to \bar qg\gamma$, $qq \to qq\gamma$, $q\bar q \to q \bar q\gamma$, $\bar q \bar q \to \bar q \bar q \gamma$. In our approach, we apply the high energy approximation for all the incoming partons. Since in this regime the gluon density is strongly dominant,  it drives evolution of the sea quark densities. Therefore, we shall keep only the contributions with the maximal number of incoming gluons, i.e. $qg \to q\gamma$ and $\bar qg \to \bar q\gamma$ for the $2\to 2$ processes and  $gg \to q\bar q\gamma$ for the $2\to 3$ processes. 

We apply the $k_T$-factorization approach for these processes in which gluons carry non-zero transverse momentum and are off-shell: $g\to g^*$. For the incoming quarks and antiquarks, we neglect the transverse momentum and we use the collinear approximation. This setup is often called a hybrid factorization approach. The main motivation for this approximation scheme is our focus on the gluon transverse momentum distribution at small and moderate~$x$.

It should be stressed that the available data on the prompt photon hadroproduction at the LHC extend from the kinematic region where the values of parton $x$ are small, $x\sim 0.002$, to larger $x>0.1$. It is well known that the high energy approximation is best motivated in the former region. It was however argued in detail in Refs.~\cite{Deak:2008ky,Motyka:2016lta} that the high energy factorization scheme may provide a good approximation of the high energy amplitudes also for moderate parton~$x$. The key argument used there is based on a detailed analysis of the values of exchanged gluon kinematics in the Sudakov decomposition. In general, the momentum~$k$ of the off-shell gluon exchanged in the $t$-channel between the proton with momentum $P_1$ and the hard interaction vertex may be written as $k=xP_1 + \beta P_2 + k_{\perp}$, where the protons' momenta $P_1$ and $P_2$ are approximated to be light-like, and $k_{\perp}$ belongs to the plane orthogonal to $P_1$ and $P_2$. Then, the quality of the high energy approximation crucially depends on the value of the Sudakov parameter $\beta$ \cite{Deak:2008ky,Motyka:2016lta}, which is typically much smaller than the value of parton~$x$. These two values are correlated because they both are inversely proportional to the collision energy. The gluon $x$, however, is driven by the large invariant mass of the state produced in the hard collision, while the Sudakov parameter $\beta$ is driven by a usually much smaller mass of the proton remnant. Hence, $\beta \ll x$, and the high energy approximation is still applicable at moderate~$x$. For  a detailed discussion of this issue and the gauge invariance problem  in this approximation scheme for related Drell--Yan and $Z^0$ hadroproduction processes, see Ref.~\cite{Motyka:2016lta}.

\subsection{Photon production in the $qg^*$ partonic channel}
\label{Sec:2.1}

We start the description of the prompt photon hadroproduction in the $k_T$-factorization approach with the $2 \to 2$ approximation, i.e.\ with the $qg^* \to q\gamma$ and $\bar qg^* \to \bar q \gamma$ channels. In what follows, we use the light cone coordinates of 
the photon polarization vectors corresponding to helicities $\sigma = \pm$:
\be\label{eq:polvec}
\epsilon_{(\sigma)} = (0,0,\vec{\epsilon}_{\perp} ^{(\sigma)}), 
\ee
where $\vec{\epsilon}_{\perp} ^{(\pm)} = \mp(1,\pm i)/\sqrt{2}$.
The quark--gluon channel contribution to the photon production with rapidity $y$ and transverse momentum $\vec{q}_T$, 
derived in \cite{Kopeliovich:1995an,Kopeliovich:1998nw,Motyka:2014lya},  reads
\begin{align}\nonumber
\frac{\tHi _{\sigma}}{d y \,d^2 \vec{q}_T}&=  \frac{4\pi \alphas (\mu_R)}{3}
\int_{x_F}^1 dx_q  \sum_{i \in \{f,\bar f\}} e_i^2\, q_{i}(x_q,\mu_F) 
 \\
& \times 
\int {d^2 \vec{k}_T \over \pi {k}_T ^2}\, F_g({x}_g,{k}_T,\mu_F) \;
\tilde{\Phi} _{\sigma \sigma} (\vec{q}_T,\vec{k}_T, x_F/x_q),
\label{dsigma2b}
\end{align}
where $k_T=|\vec{k}_T|$, $e_i$ are quark charges in units of the elementary charge $e$,  $q_i$ are collinear distributions of quarks and antiquarks with flavors $f$ and 
$\bar{f}$ (not multiplied by $x$). In addition,  the photon Feynman variable reads
\begin{equation}
 x_F=\frac{q_T}{\sqrt{S}} \,e^y ,
  \label{xF_rapidity}
\end{equation}
while the  gluon momentum fraction $x_g$ is obtained from kinematics of the parton scattering in the high energy limit
\begin{equation}
x_g= \frac{q_T}{\sqrt{S}}\,e^{-y}+ \frac{z(\vec{k}_T- \vec{q}_T)^2 }{ (1-z) q_T\sqrt{S} }\,e^{-y},
\label{x_g_gamma}
\end{equation}
where $z=x_F/x_q$. From now on, we keep the renormalization and factorization scales to be equal, $\mu_R=\mu_F$.

The function $F_g$ is the transverse momentum dependent (or unintegrated) gluon density, in the form fixed by the leading logarithmic relation to the collinear gluon density $g(x,\mu_F)$: 
\be
\int_0^{\mu_F ^2} d{k}_T ^2\, F_g(x,{k}_T,\mu_F) = xg(x,\mu_F).
\ee
All quarks are assumed to be massless when compared to the photon transverse momentum $q_T$. The diagonal 
in photon helicity impact factors are given by
\begin{eqnarray}
\tilde{\Phi}_{\sigma \sigma} (\vec{q}_T,\vec{k}_T,z)= \sum_{\lambda_1,\lambda_2\in\{+,-\}} \ {A_{\lambda_1,\lambda_2}^{(\sigma)}}^\dagger A_{\lambda_1,\lambda_2}^{(\sigma)}\, ,
\label{formfactor}
\end{eqnarray}
where $\lambda_1$ and $\lambda_2$ are helicities of the incoming and outgoing quark, respectively.
With the chosen  photon polarization vectors (\ref{eq:polvec}), the functions
\be
A_{\lambda_1,\lambda_2}^{(\pm)} 
= \frac{e}{8\pi} \,
\delta_{\lambda_1,\lambda_2}\,(2-z \mp \lambda_1 z) 
\left[ \frac{- \vec{q}_T }{\vec{q}_T^{\ 2}} - \frac{-(\vec{q}_T-z\vec{k}_T) }{(\vec{q}_T-z\vec{k}_T)^2 }  \right] \cdot \vec{\epsilon}_\bot^{\ (\pm)} 
\label{amplitudesMom+}
\ee
are proportional to the photon emission amplitudes. 
After summation over quark and photon helicities, the cross section for the 
real photon hadroproduction reads
\begin{align}\nonumber
\frac{d\sigma^\gamma}{d y d ^2\vec{q}_T} &= \frac{\alpha_{\mathrm{em}}}{3\pi}  
\int_{x_F}^1 \frac{dz}{z} \,      \frac{x_F}{z}\, \sum_{i\in\{f,\bar{f}\}}e_i^2\,q_i\!\left(\frac{x_F}{z},\mu_F\right)
\\
&\times
 \int \frac{d^2 \vec{k}_T}{k_T^2} \alpha_s F_g({x}_g,k_T,\mu_F)  	\,
\frac{\left[ 1+ (1-z)^2\right] z^2 \vec{k}_T^2 }{\vec{q}_T^{\ 2} (\vec{q}_T-z\vec{k}_T)^2} \, + (y \rightarrow -y), 
\label{dsigma7}
\end{align}
where the symmetrization $y \rightarrow -y$ since both the  initial states, $q(k_1)\,g^*(k_2)$ and  $g^*(k_1)\,q(k_2)$, contribute.  

\begin{figure}[t]
\centering
\begin{tabular}{ll}
\includegraphics[width=.37\textwidth]{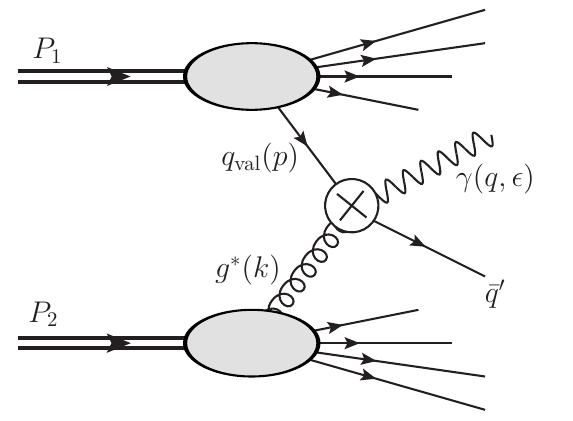} \hspace{1em} &
\includegraphics[width=.37\textwidth]{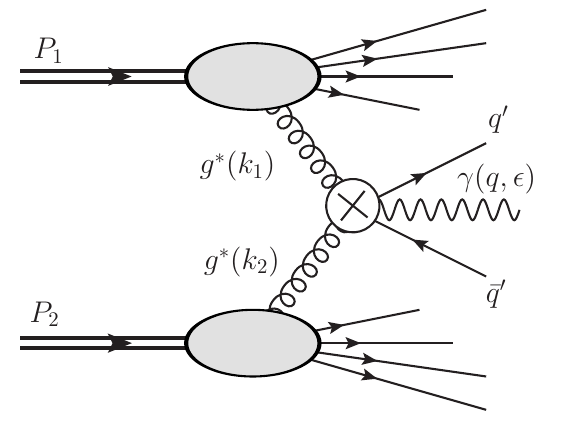} \\
{\large a)} & {\large b)} 
\end{tabular}
\caption{Partonic channels in $\gamma$ hadroproduction: 
a) the $\qval g^* \to q\gamma$ channel and b) the $g^* g^* \to q\bar q \gamma$ channel. The crossed blobs represent channel dependent hard scattering amplitudes. 
\label{Fig:channels}}
\end{figure}

The integrand in (\ref{dsigma7}) is singular  for $\vec{q}_T = z\vec{k}_T$. This is a collinear singularity coming from the emitted photon momentum parallel to the outgoing quark momentum. In experiments, however, the photon measurement requires a separation from the outgoing hadrons (or jets). Hence, the experimental data assume certain isolation cut,  $\Theta_{R_{\gamma}}(q_{\gamma},\{p^H _i\})$,
which depends on the photon momentum $q_\gamma$, the hadron momenta $\{p^H _i\}$ and the isolation cone parameter $R_{\gamma}$. In the parton level formulas, the isolation criterion is implemented with outgoing parton momenta  $\{p_i\}$ instead of the hadronic momenta $\{p^H _i\}$. Thus, in order to obtain a physically meaningful result,  Eq.~(\ref{dsigma7}) must be complemented by a suitable cut
$\Theta_{R_{\gamma}}(\vec{q}_{T},z,\vec{k}_T)$. The detailed discussion of our implementation  is given section  (\ref{sec:cut}).

The $qg^* \to q\gamma$ channel described in the high energy limit within the $k_T$-factorization framework was shown \cite{Kopeliovich:1995an,Kopeliovich:1998nw} to be equivalent to the dipole picture of high energy scattering, in close analogy to the dipole picture of the Drell--Yan process \cite{Kopeliovich:1995an,Kopeliovich:1998nw,Brodsky:1996nj}, see also \cite{Motyka:2014lya,Brzeminski:2016lwh} for recent developments. This picture is obtained by the Fourier transform of the impact factors to the impact parameter space, and the color dipole scattering amplitude emerges as a result of an interference between the initial and final state photon emission amplitudes. This approach has an advantage to be capable to efficiently include the effects of multiple scattering and it was explored for instance in recent analyses \cite{Santos:2020nqy,Goncalves:2020tvh}.

\subsection{Photon production in the $g^*g^*$ partonic channel}
\label{Sec:2.2}

In an alternative approach to the prompt photon production in the high energy approximation, one generates  sea quarks from the gluon in the last splitting. With this assumption, the sea quark contributions $q_{\rm sea}$ can be absorbed into the $g^* g^* \to q\bar q \gamma$ hard matrix elements. 
The valence quark  contribution $\qval$, however, is not included in the $g^*g^*$ channel, and these two contributions enter the cross section additively,
see Fig.\ \ref{Fig:channels},
\be
d\sigma^{\gamma}  = d\sigma^{(\qval g^* \to q\gamma)} +  d\sigma^{(g^*g^* \to q\bar q\gamma)} .
\ee

The  $g^*g^* \to q\bar{q} \gamma$ 
hard subprocesses cross section is calculated in the $k_T$-factorization framework. The scheme and details of the calculations follow closely more general calculations performed in the same setup for the Drell--Yan structure functions \cite{Motyka:2016lta}. In fact, the formulas for the photon production may be recovered from those derived in Ref.\  \cite{Motyka:2016lta} by removing leptonic part with boson propagator, taking the limit $M \to 0$ for the 
 Drell--Yan intermediate boson mass and considering only the diagonal in helicity structure functions for the transverse polarizations. For completeness, we shortly repeat the main steps of these calculations adjusted to the photon hadroproduction case.  
 
 \begin{figure}[t]
\centering
\includegraphics[width=.18\textwidth]{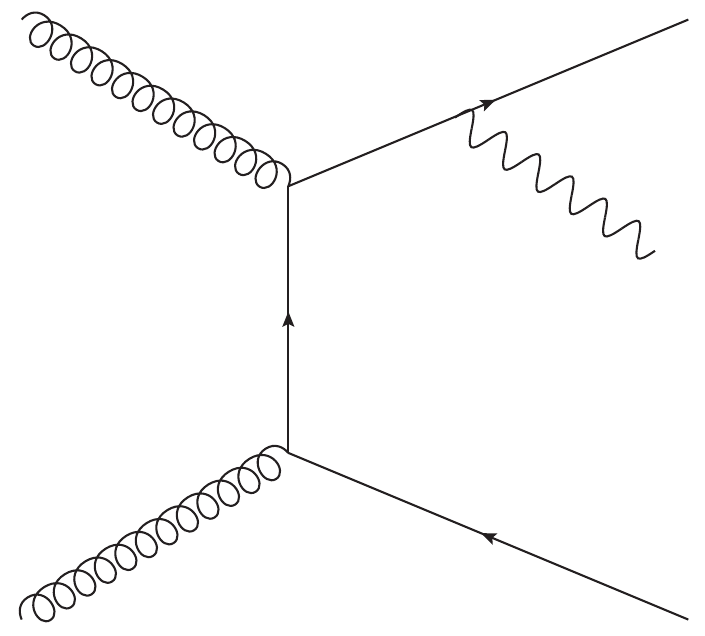} \hspace{5em}
\includegraphics[width=.18\textwidth]{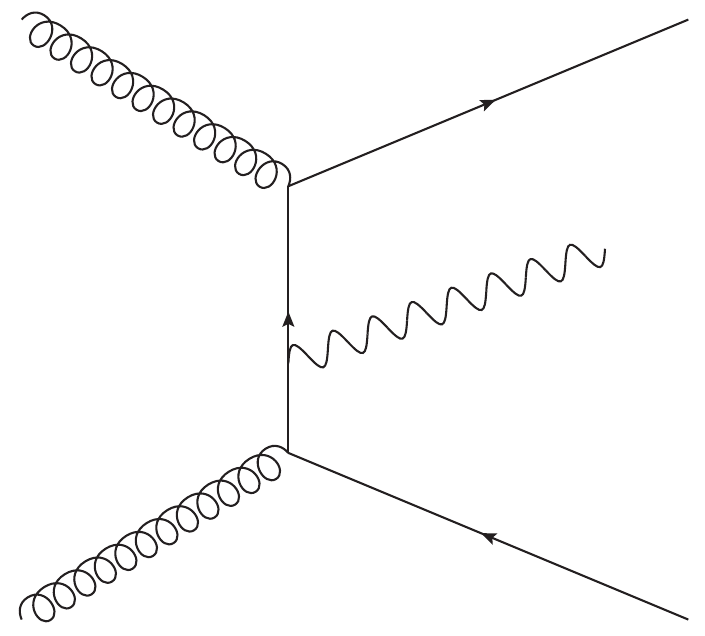} \hspace{5em}
\includegraphics[width=.18\textwidth]{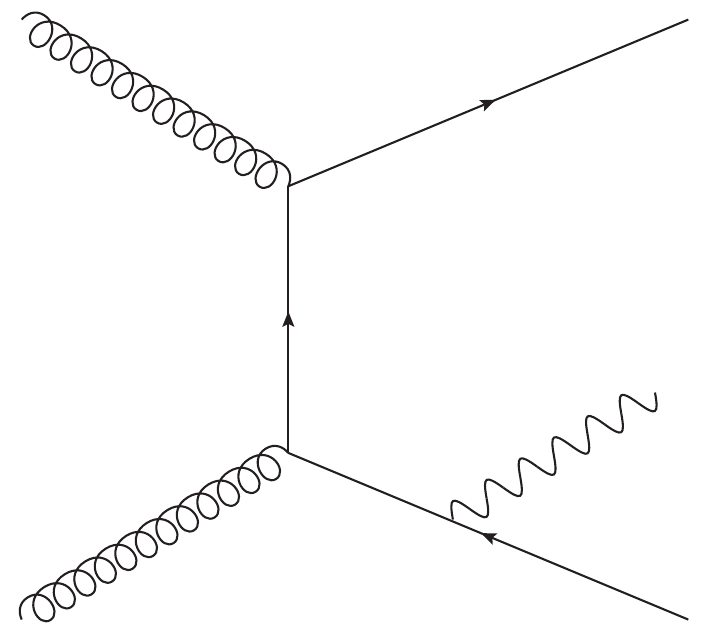} \\[1em] 
\includegraphics[width=.18\textwidth]{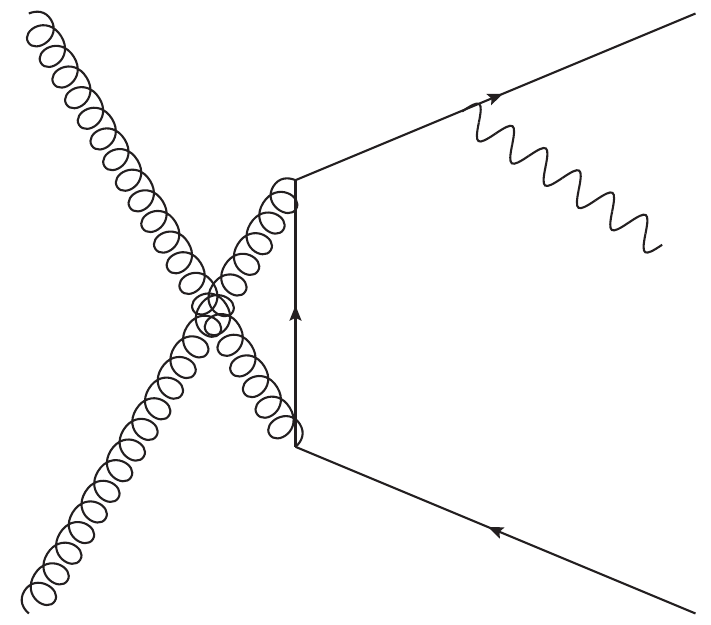} \hspace{5em}
\includegraphics[width=.18\textwidth]{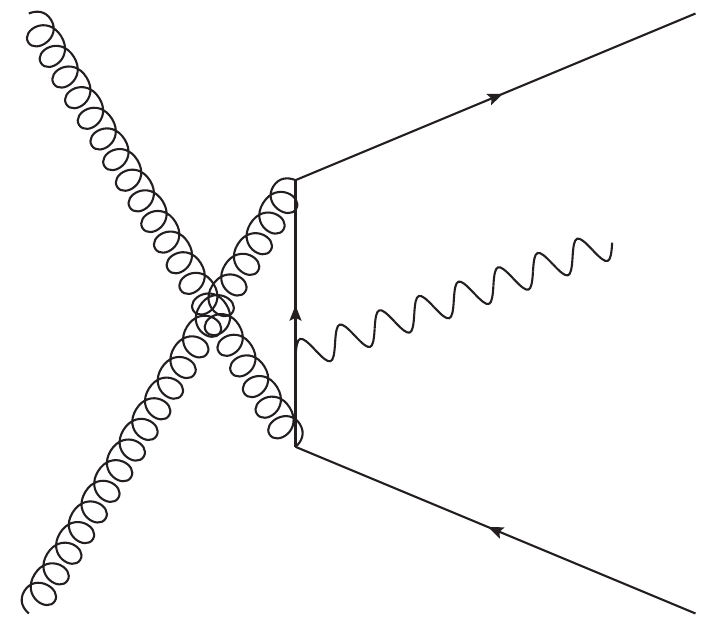} \hspace{5em} 
\includegraphics[width=.18\textwidth]{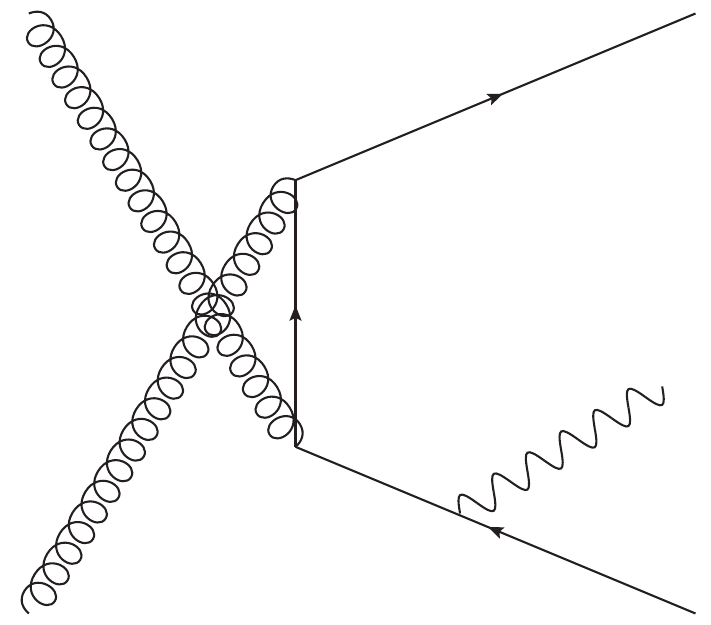} \\[1em]  
\includegraphics[width=.18\textwidth]{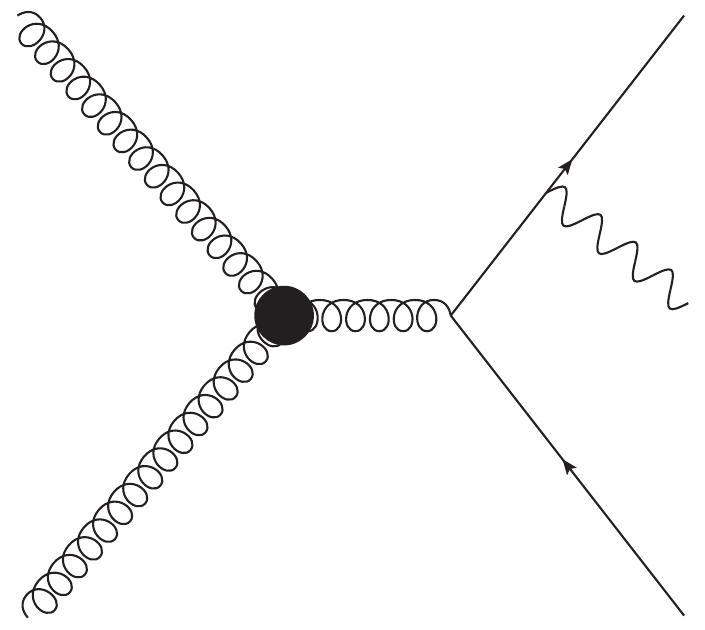} \hspace{5em} 
\includegraphics[width=.18\textwidth]{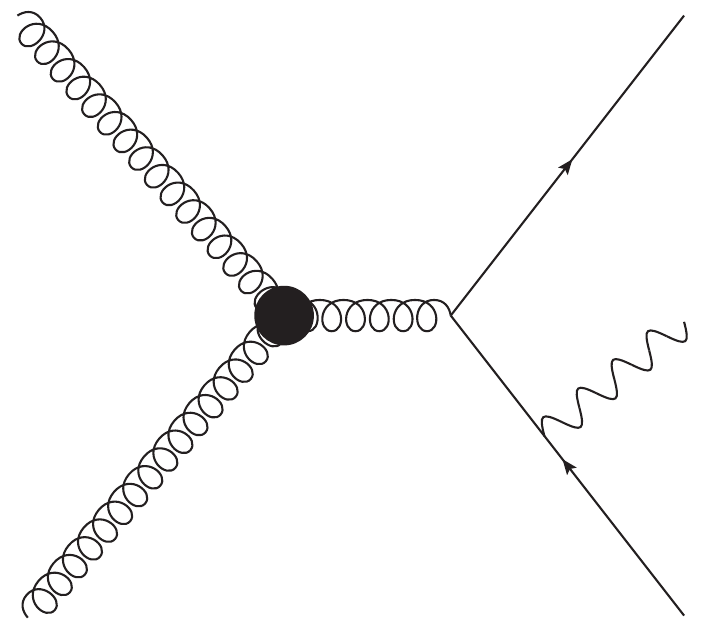} \\ 
\caption{Eight Feynman diagrams that contribute to the $g^* g^* \to q\bar q \gamma$ partonic channel corresponding to the amplitudes ${\cal M}^{(i)}$ for $i=1,2,\ldots,8$, respectively. The black blobs  denote the effective triple gluon vertex $V_{\mathrm{eff}}$, see Appendix of Ref.~\cite{Motyka:2016lta} for  its  definition. 
\label{Fig:ggdiags}}
\end{figure}

In the $k_T$-factorization gluons are virtual, $k_i^2 \simeq -\vec{k}_{iT} ^2 <0$, and  quarks are taken to be massless, $p^2 _3 = p_4 ^4 = 0$. The standard gluon momenta decomposition in the high energy limit is applied: $k_1 = x_1 P_1 + k_{1\,\perp}$ and $k_2 = x_2 P_2 + k_{2\,\perp}$, see also the discussion in Sec.~\ref{Sec:2}, preceding Sec.~\ref{Sec:2.1}. The parton level scattering amplitude is a sum of eight diagrams shown in Fig.\ \ref{Fig:ggdiags}. The high energy limit for virtual gluon polarizations is used, in which the virtual gluon polarization vectors $\pi_{g^*}(k_i)$ are approximated by the so-called ``nonsense polarizations'', using the so-called Collins--Ellis trick \cite{Collins:1991ty} in the derivations,
\be
\pi^{\mu} _{g^*}(k_1) \simeq x_1 P^{\mu} _1 / \sqrt{\vec{k}_{1T} ^2}\,,~~~~~~~~~~~~\pi^{\mu} _{g^*}(k_2) \simeq x_2 P^{\mu} _2 / \sqrt{\vec{k}_{2T} ^2}\,.
\ee
The impact factors ${\cal T}_{\mu} ^{(i)}$ with $i=1,2,\ldots,8$ are defined as
\be
{\cal T}_{\mu} ^{(i)} = {\cal M}^{(i)} _{\mu, \alpha\beta} \,P_1 ^{\alpha} P_2 ^{\beta},
\ee
where ${\cal M}^{(i)} _{\mu, \alpha\beta}$ are the amplitudes for the diagrams shown in Fig.\ \ref{Fig:ggdiags} with amputated polarization vectors of the incoming gluons and  outgoing photon. The explicit expressions for the impact factors are given in Appendix of Ref. \cite{Motyka:2016lta} . 

The $g^* g^* \to q\bar q \gamma$ impact factor, given by ${\cal T}^{g^*g^*} _{\mu} = \sum_{i=1} ^8 {\cal T}^{(i)} _{\mu}$,   is used to calculate the $g^*g^*$ channel contributions cross sections for photon production with helicity $\sigma$, 
\begin{align}\nonumber
d\sigma^{(g^*g^* \to q\bar q \gamma)}  _{\sigma}& = \int dx_1 \int {d^2 \vec{k}_{1T} \over \pi {k}_{1T}^2}  F_g(x_1,{k}_{1T},\mu_F) \int dx_2 \int {d^2 \vec{k}_{2T}
 \over \pi {k}_{2T} ^2 } F_g(x_2,{k}_{2T},\mu_F) 
\\
& \times \; {(2\pi)^4\over 2\sp}\,  {\cal H}_{\sigma} \,dP S_3(k_1 + k_2 \to p_3 + p_4 + q), 
\end{align}
with $F_g$ being the unintegrated gluon distribution introduced in the previous section and 
\be
{\cal H}_{\sigma}  = \sum_{f\in\{u,d,s,c,b\}}{1 \over (N_c^2 - 1)^2}\sum_{a,b}\sum_{i_3,i_4}\sum_{r_3,r_4}
 \left({\cal T}^{g^*g^*}_{\mu} \epsilon^{\mu} _{(\sigma)}\right) \left({\cal T}^{g^*g^*}_{\nu} \epsilon^{\nu} _{(\sigma)}\right)^\dagger ,
\ee
where summations are performed over quark flavors $f$ (present through the charge $e_f$ in the amplitudes ${\cal T}^{g^*g^*}_{\mu}$),  color indices  of the gluons $(a,b)$ and   quarks $(i_3,i_4)$, and over the quark helicities 
$(r_3,r_4)$. The latter summation leads to traces over Dirac spinors which are evaluated with the FORM program for symbolic manipulations 
\cite{Vermaseren:2000nd}. 
The resulting expressions, obtained in two independent calculations, 
are lengthy and do not need to be explicitly displayed. It was also  checked that ${\cal H}_{\sigma}\sim {k}_{iT} ^2 $ 
when the gluon transverse momentum ${k}_{iT} ^2 \to 0$, as required by the gauge invariance condition in the high energy limit.

The phase space for the final state particles of partonic scattering is parametrized in terms of the photon variables $(y,\vec{q}_T)$
and the variables $(z,\phi_{\kappa})$ describing the $q\bar q$ kinematic configuration,   
\be
dPS_3(k_1 + k_2 \to p_3 + p_4 + q) = {dy \, d^2 \vec{q}_T \, dz \, d\phi_{\kappa} \over 8(2\pi)^9}d\kappa^2 \ee
\[
\times\, \delta\! \left[ \kappa^2 -   z(1-z)\!\left ( x_{q\bar q}x_2 S -x_{q\bar q}\frac{q_T^2 }{x_F} -  \vec{\Delta}^2\right)\right],
\]
where the variables $(z,\vec \kappa)$ are implicitly defined by the parametrization of the quark and antiquark momenta
\be
p_3 = z x_{q\bar q} P_1 + {\vec{p}_3 ^2 \over z x_{q\bar q} S } P_2  +  p_{3\perp}, \qquad  
p_4 = (1-z) x_{q\bar q} P_1 + {\vec{p}_4 ^2 \over (1-z) x_{q\bar q} S } P_2  +  p_{4\perp},
\ee
in which
\begin{align}\nonumber
p_{3\perp} = (0,0,\vec{p}_3),\quad p_{4\perp} &= (0,0,\vec{p}_4), \quad 
\vec{p}_3 = z \vec{\Delta} + \vec{\kappa}, \quad
\vec{p}_4 = (1-z) \vec{\Delta} - \vec{\kappa}, 
\\
\vec{\Delta} &= \vec{k}_1 + \vec{k}_2 - \vec{q}, \quad x_{q\bar q} = x_1 - x_F.
\end{align}
For the comparison with data it is necessary to integrate over the final
state quark/antiquark kinematical variables and sum over the photon polarizations, 
\be
{d\sigma^{ \gamma} \over dy d^2 \vec{q}_T} = 
\int dz \int d\phi_{\kappa}  \,   \sum_{\sigma=\pm}{ d\sigma^{(g^*g^* \to q\bar q \gamma)}  _{\sigma} \over  dy d^2 \vec{q}_T dz d\phi_{\kappa}}.
\ee
\vskip 4mm

\section{Transverse momentum dependent gluon distributions}
\label{Sec:3} 

Several parametrizations  of the transverse momentum dependent gluon distribution in the proton, $F_g(x,{k}_T,\mu_F)$, were proposed. Many of them were derived in the regime of small-$x$ dynamics, including solutions of the BFKL equation~\cite{Fadin:1975cb,Kuraev:1977fs,Balitsky:1978ic} or the Balitsky--Kovchegov (BK)~\cite{Balitsky:1995ub,Kovchegov:1999yj, Kovchegov:1999ua} equation. The data for the prompt photon production, however, extend out of the small-$x$ domain. Therefore, we consider parametrizations that may be used also for moderate values of $x>0.01$. 
The most widely used approaches having this feature are  the following:
\begin{itemize}
\item[A.]
The Kimber--Martin--Ryskin (KMR) approach \cite{Kimber:1999xc,Kimber:2001sc} that permits to recover the unintegrated parton distributions from the collinear parton distribution functions.
\item[B.]
The approach based on the solutions of the CCFM equation \cite{Ciafaloni:1987ur, Catani:1989sg,Catani:1989yc,Marchesini:1994wr}, implemented by  Jung and Hansson  (JH) \cite{Hansson:2003xz} and later on by Jung and Hautmann (JH-2013)  \cite{Hautmann:2013tba}.
\item[C.]
 We also consider the gluon distribution from the Golec-Biernat--W\"{u}sthoff (GBW) model \cite{GolecBiernat:1998js}, extended phenomenologically to the values of $x>0.01$. 
 \end{itemize} 
 All of these parametrizations were optimized to  describe the HERA data on the proton structure function $F_2$ (for the GBW gluon distribution, the optimization was performed for $x\le 0.01$ but not for its phenomenological extension).

\subsection{KMR-AO gluon distribution}
The KMR scheme was used in several variations. Our choice is the integral version of the KMR prescription \cite{Golec-Biernat:2018hqo} in which
the unintegrated gluon distribution for $k_T> Q_0=1~{\rm GeV}$ is given by
\be
\label{eq:3.1}
 F_g(x,\kperp,Q) \equiv 
\frac{T_a(Q,\kperp)}{\kperp^2}\sum_{\aprime\in\{f,\bar{f},g\}}
\int_x^{1-\Delta(\kperp,Q)} \frac{dz}{z}\,P_{g\aprime}(z,\kperp)\, 
D_{\aprime}\Big(\frac{x}{z},\kperp\Big) \, ,
\ee
where the function $T_a$ is called the Sudakov form factor
\be
T_g(Q,\kperp) =\exp\left\{
-\int_{\kdwaperp}^{Q^2} \frac{d\pperp^2}{\pperp^2}\sum_{\aprime\in\{f,\bar{f},g\}}\int_0^{1-\Delta(\pperp,Q)} dz z P_{\aprime g}(z,\pperp)
\right\}\, .
\ee
In the above, $f/\bar{f}$ denote quark/antiquark flavors,  $P_{aa^\prime}(z,\mu)$ are the Altarelli--Parisi splitting functions given in terms of the expansion in 
$\alpha_s(\mu)/(2\pi)$ and  $D_a(x,\mu)$ are collinear parton distribution functions (PDFs). We choose the leading order  splitting functions and the CT10 PDFs
\cite{Lai:2010vv}. We consider the angular ordering version of the KMR distribution (KMR-AO) in which the function $\Delta$ in the upper integration limits equals
\be
\Delta(\kperp,Q)=\frac{\kperp}{\kperp+Q}\, .
\ee
This prescription  imposes angular ordering in the last step of the evolution \cite{Kimber:2001sc}. 
 For the values $k_T\le Q_0$, the unintegrated gluon distribution is frozen at the boundary value $F_g(x,Q_0,Q)$.

We prefer to use the integral version of the KMR distribution over the differential one since it allows to avoid problems for large values of  transverse momentum, $k_T>Q$, discussed in \cite{Golec-Biernat:2018hqo}.

\subsection{JH and JH-2013 gluon distribution} 

The unintegrated gluon distribution in the all-loop CCFM scheme  \cite{Marchesini:1994wr,Hautmann:2017fcj,Golec-Biernat:2019scr} takes into account small-$x$ coherence effects which are reflected in
angular ordering in gluon cascade branching. This leads to a non-Sudakov form factor which screens the $1/z$ singularity in the $P_{gg}$ splitting function.
The CCFM scheme was extended for gluons to the region of large $x$ by taking into account terms with finite $z$  in $P_{gg}$, and the Sudakov form factor  with  angular ordering of the final state emissions. Infrared parameters in such a scheme were fitted to the HERA data on $F_2$ by Jung and Hansson   in \cite{Hansson:2003xz}, which analysis offers the JH unintegrated gluon distribution. Along similar lines in the year 2013 a new parametrization (called
JH-2013) of the unintegrated gluon distribution was constructed by Jung and Hautmann \cite{Hautmann:2013tba}.
The new parametrization is based on the angular ordered CCFM gluon branching scheme with unintegrated valence quark distributions included.


\subsection{Extended GBW gluon distribution} 

The  unintegrated gluon density $F_g$ from the GBW saturation model is given by
\begin{equation}
\alpha_s F_g(x,k_T,\mu)  = \frac{3\sigma_0}{4\pi^2} \frac{k_T^2}{Q_s^2} \exp(-k_T^2/Q_s^2) \times \left(\frac{1-x} {1-0.01}\right)^7\; ,
\label{GBWgluon}
\end{equation}
where the the saturation scale $Q^2_s=(x/x_0)^{-\lambda}~ {\rm GeV}^2$. We use the  parameters from the recent refit of the GBW  model 
done in \cite{Golec-Biernat:2017lfv}:  $\sigma_0=27.32~{\rm mb}$, $x_0 = 0.42\cdot 10^{-4}$ and $\lambda=0.248$. 
 Since the GBW model was fitted to the  data  with $x<0.01$,  the original form of the gluon distribution was   extrapolated to the values of  $x>0.01$
 by multiplying by the factor $(1-x)^7/(1-0.01)^7$, that ensures a smooth transition from the small-$x$ domain to the region  of $x\sim 1$. 
 The form of the distribution for $x\to 1$ follows from dimensional scaling rules of the Regge formalism close to the kinematic end point. The main purpose of using the extrapolated GBW gluon distribution is its distinct transverse momentum dependence that is exponentially suppressed for $k_T > 1~{\rm GeV}$. 
 Such a  strong and narrow in $k_T$ suppression is not present in the KMR-AO and JH gluon distributions.

\subsection{Gluon distribution comparison}

\begin{figure}[t]
\begin{center}
\includegraphics[width=0.49\textwidth]{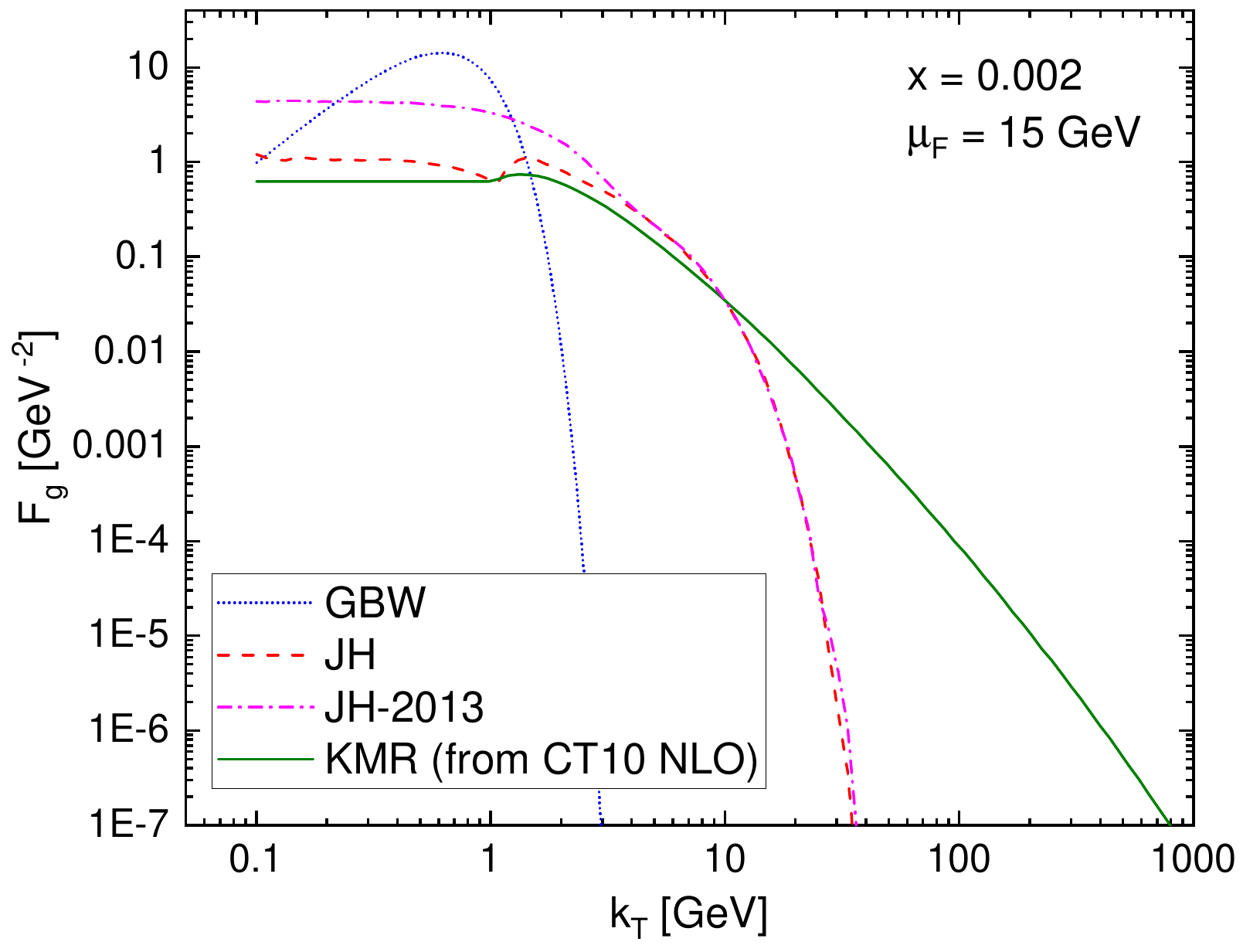}
\includegraphics[width=0.49\textwidth]{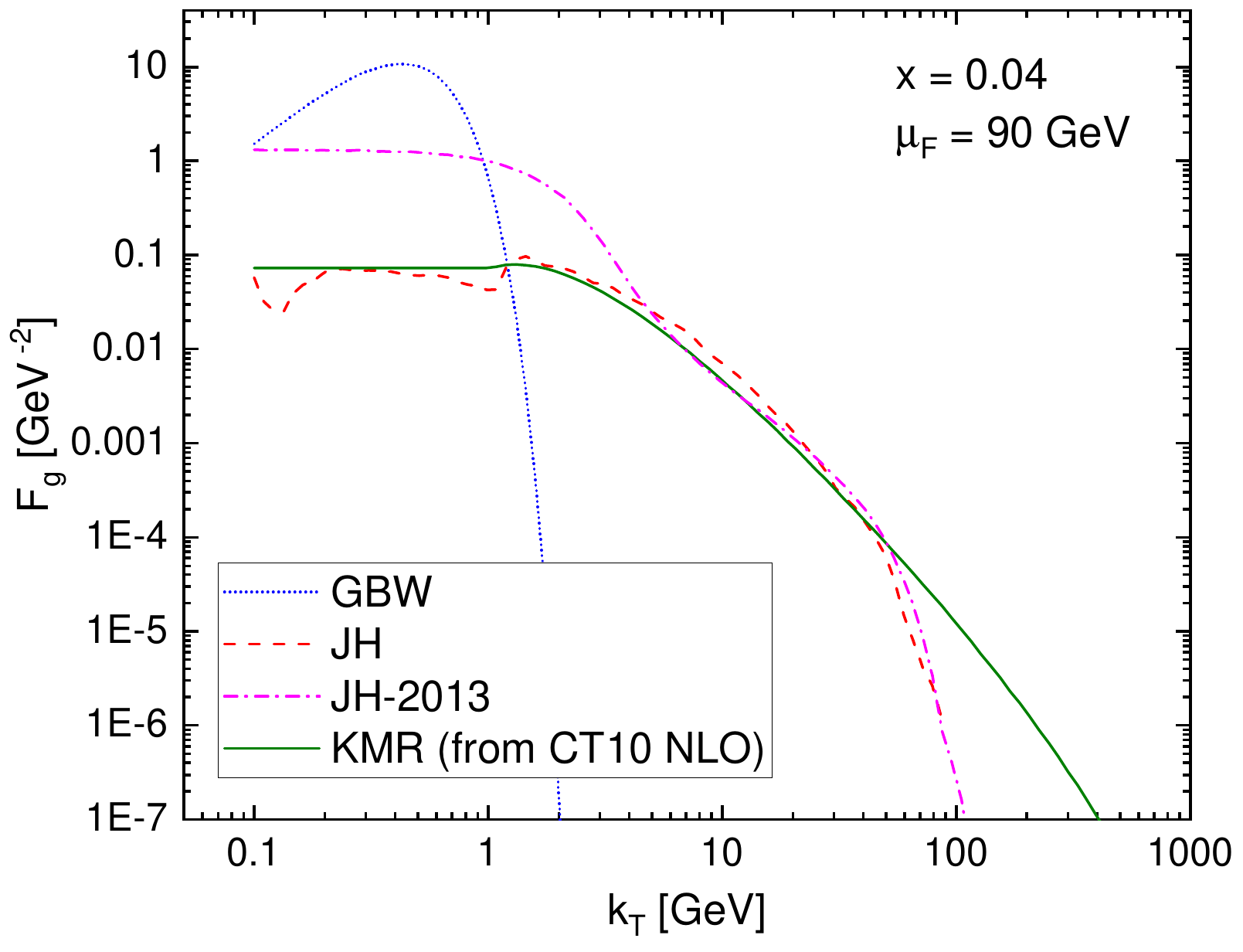}
\end{center}
\vskip -5mm
\caption{Four  unintegrated gluon distributions $F_g(x,\kperp,\mu)$ (see the text) as a function of gluon transverse momentum $k_T$ for two indicated
values of gluon $x$ and factorization scales $\mu_F$.
In the computation of the  GBW gluon distribution,  the leading order running $\alpha_s=\alpha_s(\mu_F) $ was used.}
\label{updf_comparison}
\end{figure}

In  Fig.~\ref{updf_comparison},  we illustrate the properties of the considered unintegrated gluon distributions, which are shown as a function of gluon transverse momentum~$k_T$ for two values of the longitudinal momentum fraction, $x=0.002$ and $0.04$, and two values of the factorization scale, $\mu_F=15$ and 90~GeV, correspondingly. They are motivated by  typical values in the prompt photon production at the LHC in the central region when $\mu_F=q_T$.   
Clearly, the GBW distribution is much narrower and strongly peaked at small~$k_T$ than other distributions,  which
are similar in the region $Q_0=1~{\rm GeV}<k_T < \mu_F$ but differ at the ends of the $k_T$ spectrum. In particular, while
the KMR and JH distributions are almost the same for $k<Q_0$, the new distribution JH-2013 is significantly higher.
On the other hand, while  the JH and JH-2013 distributions are practically the same for $k_T\gg \mu_F$, they are much more strongly suppressed than the KMR distribution which decreases much slower, approximately as a moderate negative power of $k_T$. These distinct  $k_T$-shapes  should have an important impact on the predictions of  the measured photon distribution in transverse momentum $q_T$.

We also compare the $x$-dependence of the gluon distributions. We choose to illustrate this dependence by evaluating the `collinear integral' of the unintegrated gluon distributions
\be\label{eq:3.2}
\widetilde{xg}(x,\mu_F^2) = \int_{0} ^{\mu_F^2} dk_T^2 \,F_g(x,k_T,\mu_F).
\ee
This integral should reproduce  the true collinear gluon distribution $xg(x,\mu_F^2)$ in the leading logarithmic approximation.
 In our approach, however, the effects beyond the leading logarithmic approximation are also present,  so differences between  $xg(x,\mu_F^2)$ and $\widetilde{xg}(x,\mu_F^2)$ should appear.

\begin{figure}[t]
\begin{center}
\includegraphics[width=0.49\textwidth]{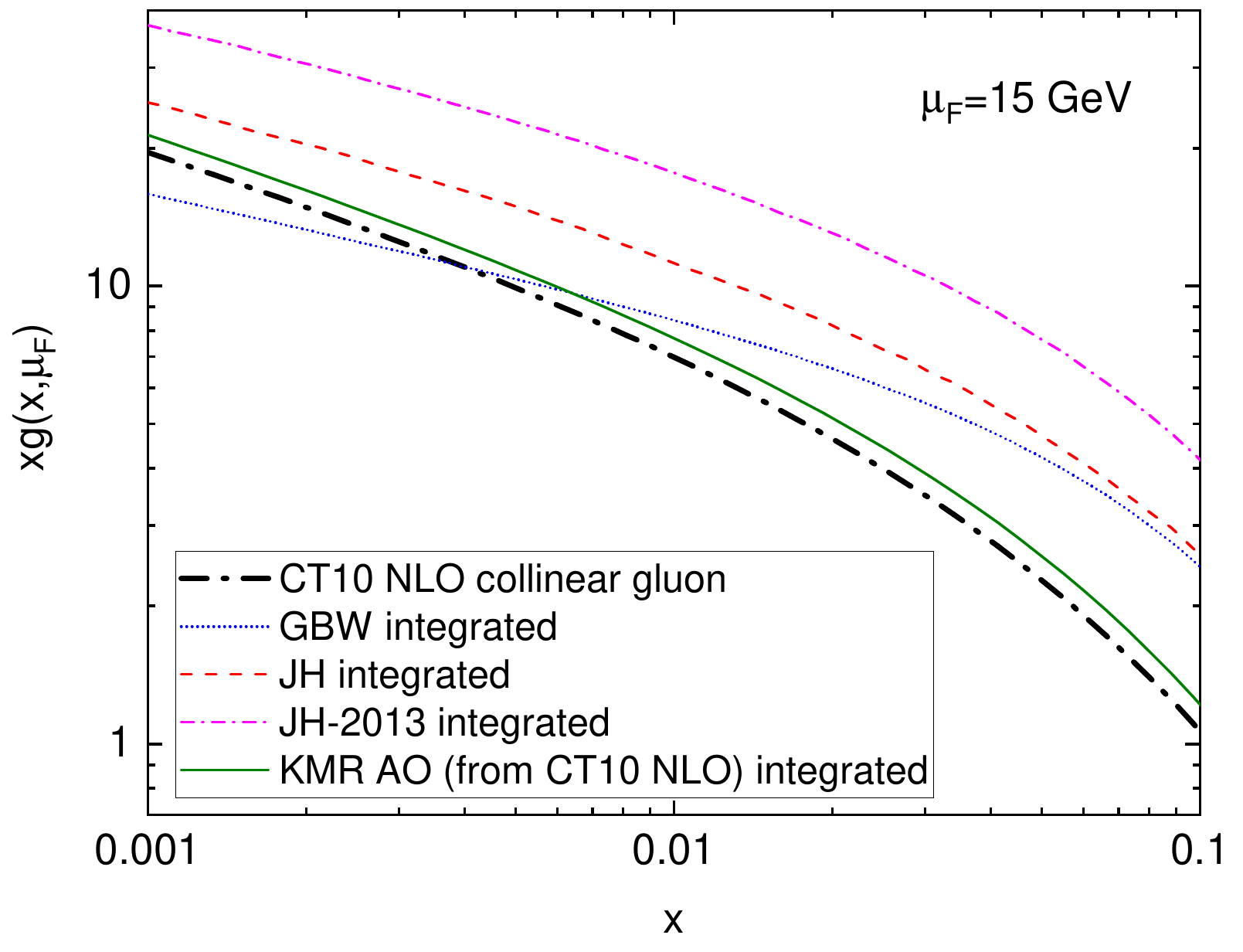}
\includegraphics[width=0.49\textwidth]{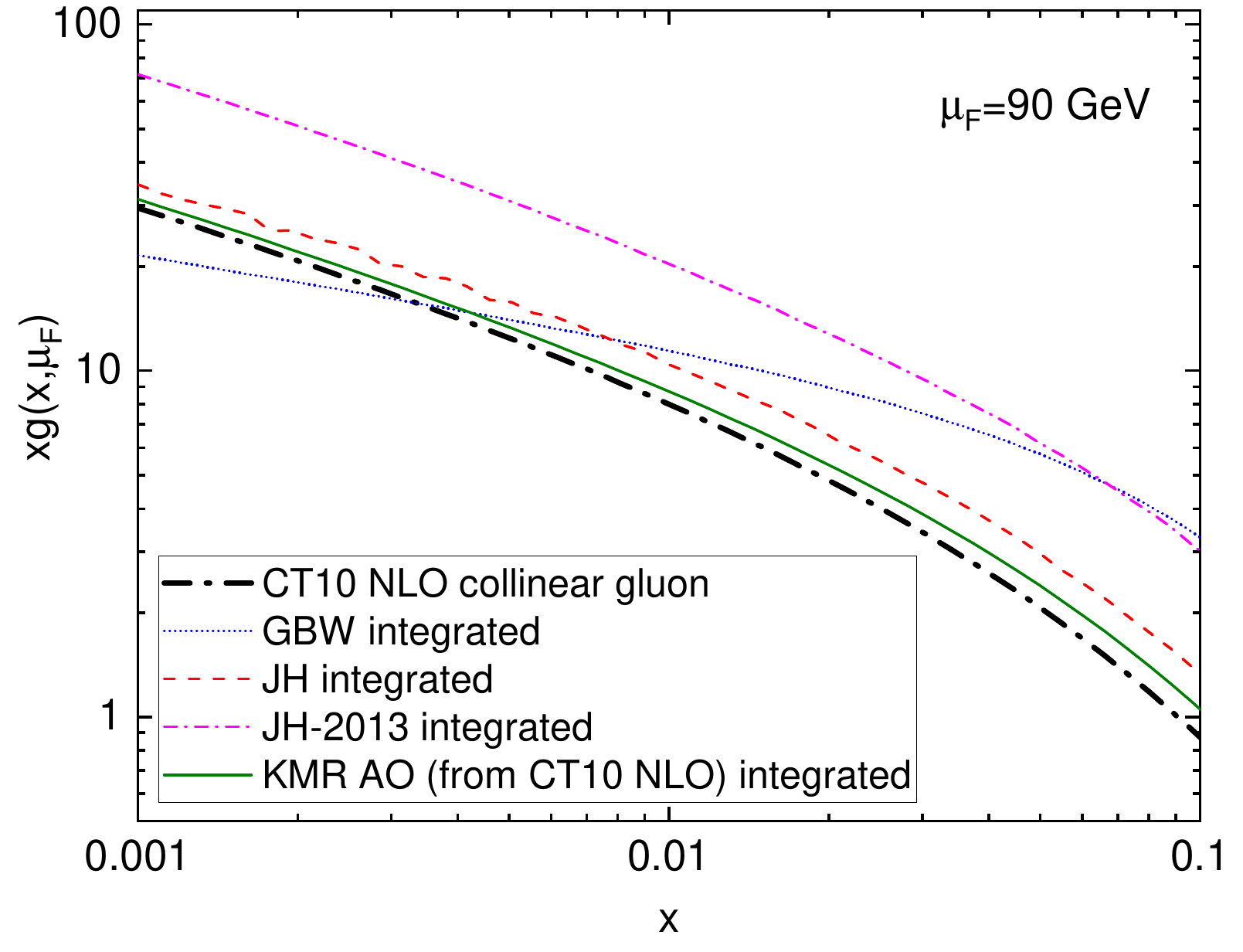}
\end{center}
\vskip -5mm
\caption{The ``collinear integral'' (\ref{eq:3.2}) computed for the four indicated unintegrated gluon distributions and two factorization scales $\mu_F$
as a function of $x$, compared to the collinear gluon distribution CT10 at the NLO.}
\label{pdf_comparison}
\end{figure}

 In Fig.~\ref{pdf_comparison},  the  gluon distribution (\ref{eq:3.2}) computed for the considered unintegrated gluon distributions
  are compared to the collinear NLO gluon distribution CT10 \cite{Lai:2010vv} from which the KMR-AO distribution is extracted. 
 The integrated KMR-AO gluon is closest to the CT10  distribution, reproducing quite well the $x$-dependence but having a sightly higher normalization.  This is due to the prescription for the value of the KMR distribution (\ref{eq:3.1}) in the non-perturbative region  of  $k_T<Q_0=1~{\rm GeV}$, where $F_g$ is frozen to the value for $k_T=Q_0$. This difference could be reduced to zero by fitting the non-perturbative value of $F_g$ but paying the price of discontinuity of $F_g$ at the boundary $k_T=Q_0$. We prefer to avoid such a situation. The integrated JH and JH-2013 distributions have  similar shapes in $x$ to
the CT10 gluon distribution, but their normalizations are significantly different. This is particularly visible for the new JH-2013 distribution, which   confirms the results shown in the original paper \cite{Hautmann:2013tba}.  The shape in $x$ of the collinear integral (\ref{eq:3.2}) with the GBW distribution  is not related to the CT10 gluon shape in any way. However, the GBW curve stays in the ballpark of the CT10 gluon values for $x<0.01$,  while  for $x>0.01$ the $x$ dependence is driven by the postulated extrapolation factor $(1-x)^7$, which is clearly too mild in this region of $x$.

\section{ Comparison to data}
\label{Sec:4}

\subsection{Isolation of $\gamma$}
\label{sec:cut}

In order to  perform the comparison of our numerical results with the LHC data, we need to perform
photon isolation from jets in similar manner as it was done by the ATLAS and CMS experiments. To this end,  we implement the Frixione cone cut at the parton level \cite{Frixione:1998jh}. For the final state $q\bar q \gamma$,  we define the distances $R_i$ between the outgoing quarks, labeled by  $i=3, 4$,  and the photon:
\begin{equation}
R_{i}= \sqrt{ (\phi_i - \phi_\gamma)^2 + (y_i - y_\gamma)^2 }\, ,
\end{equation}
where $\phi_i$ are  parton azimuthal angles and $y_i$ are their rapidity. We take $R_0=0.4$. The Frixione cone isolation procedure leads then to 
the following conditions which have to  be satisfied by the kinematic configuration:
\begin{itemize}
\item if $R_3< R_4< R_0$ then
\be\label{cond_1}
p_{3T}+p_{4T}  < q_T \frac{1-\cos R_4}{1-\cos R_0}\,~~~~~~{\rm and}~~~~~~p_{3T}< q_T \frac{1-\cos R_3}{1-\cos R_0}\, ,~~~~~~~~
\ee
\item if  $ R_3 < R_0  <R_4$ then
\begin{equation}
p_{3T} < q_T \frac{1-\cos R_3}{1-\cos R_0}\, ,
\label{cond_2}
\end{equation}
\item  no constrains when $ R_0< R_3  <R_4$,
\item if $R_4> R_3$ then conditions ($\ref{cond_1}$) and ($\ref{cond_2}$) are applied with the indices $3 \leftrightarrow 4$,
\item for  simpler partonic final states,  $q\gamma$ and $\bar q \gamma$,   only condition  (\ref{cond_2}) is imposed.
\end{itemize}

\subsection{LHC data sets}

In the comparison,  we focus on the available sets of data that probe the smallest values of gluon~$x$ variable.  
Hence we choose the following measurements of the prompt photon production from the ATLAS and CMS experiments:
\begin{itemize} 
\item ATLAS data at $\sqrt{S} =7~{\rm TeV}$ with $15~{\rm GeV}<q_T<100~{\rm GeV}$ (ATLAS$@7{\rm TeV}$) \cite{Aad:2010sp},
\item CMS data at $\sqrt{S} = 7~{\rm TeV}$ with $25~{\rm GeV}<q_T<400~{\rm GeV}$ (CMS$@7{\rm TeV}$)  \cite{Chatrchyan:2011ue},
\item  ATLAS data at $\sqrt{S} =8~{\rm TeV}$ with $25~{\rm GeV}<q_T<1500~{\rm GeV}$ (ATLAS$@8{\rm TeV}$) \cite{Aad:2016xcr},
\end{itemize}
where $q_T$ is the photon transverse momentum.
 In these experiments the cross sections were measured as a function of $q_T$ in several rapidity intervals. 
 In the following figures, the experimental data are compared to theoretical predictions obtained within the $k_T$-factorization formalism in the two schemes described in the previous sections:  the $qg^*$ scheme based on the $qg^*\to q\gamma$ and  $\bar qg^*\to \bar q\gamma$ partonic channels, and  the $g^*g^*$ scheme based on a combination of the $q_{\mathrm{val}}g^*\to q\gamma$  and  
 $g^* g^* \to q\bar q \gamma$ channels. For each theoretical scheme, we display the results obtained with the three unintegrated gluon distributions: KMR-AO, JH and GBW.

\begin{figure}
\begin{center}
\includegraphics[width=0.65\textwidth]{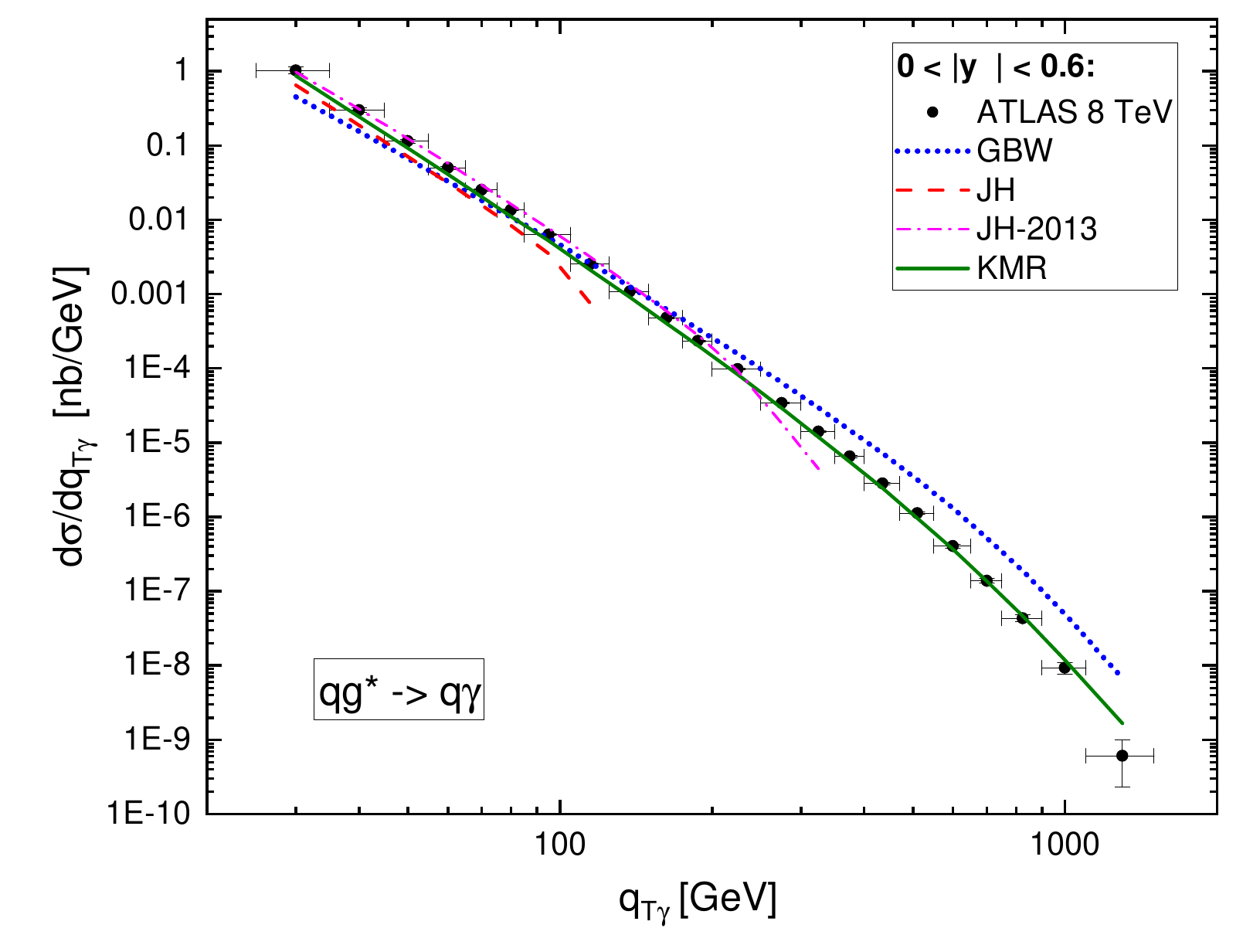} \\
\includegraphics[width=0.65\textwidth]{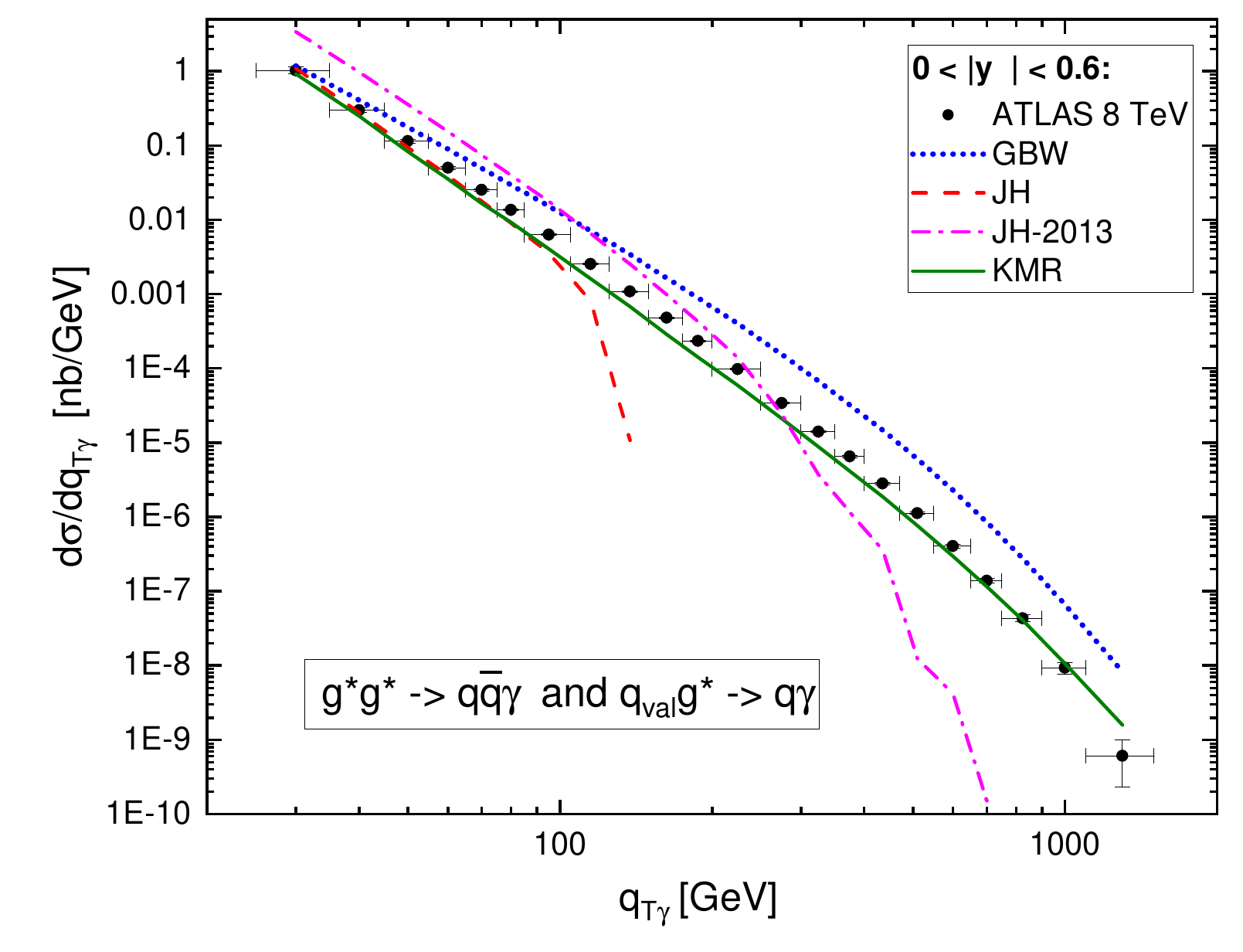}
\end{center}
\vskip -5mm
\caption{Prompt photon production cross section computed in the $qg^*$ (top) and  $g^*g^*$ (bottom) schemes with the indicated unintegrated gluon distributions,  compared to the ATLAS$@8{\rm TeV}$ data \cite{Aad:2016xcr} for the central rapidity
bin.}
\label{fig:comp_ATLAS_8}
\end{figure}

\subsection{Comparison to data}

 \begin{figure}
\begin{tabular}{cc}
\hskip -3mm
\includegraphics[width=0.45\textwidth]{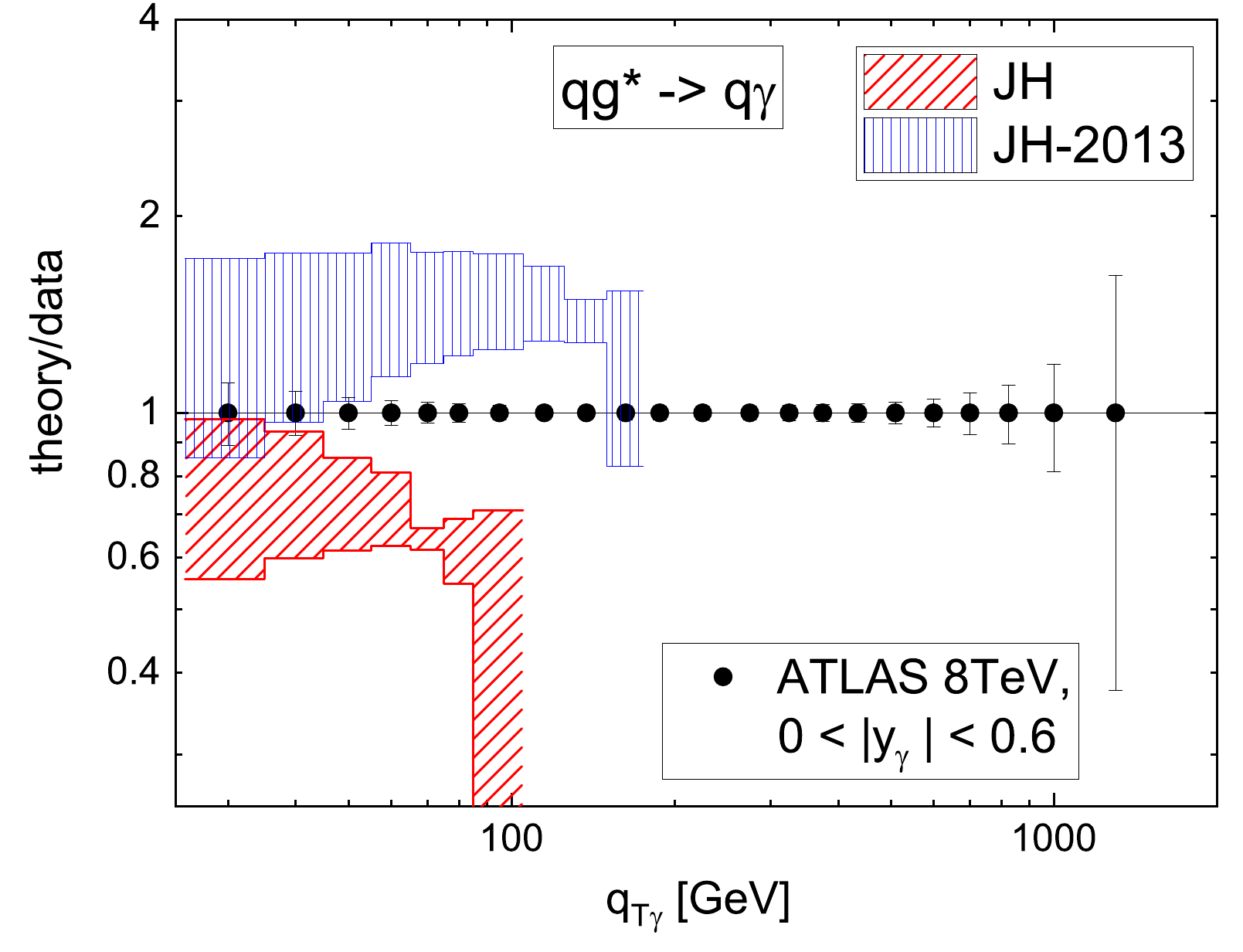}
\hskip -3mm
\includegraphics[width=0.45\textwidth]{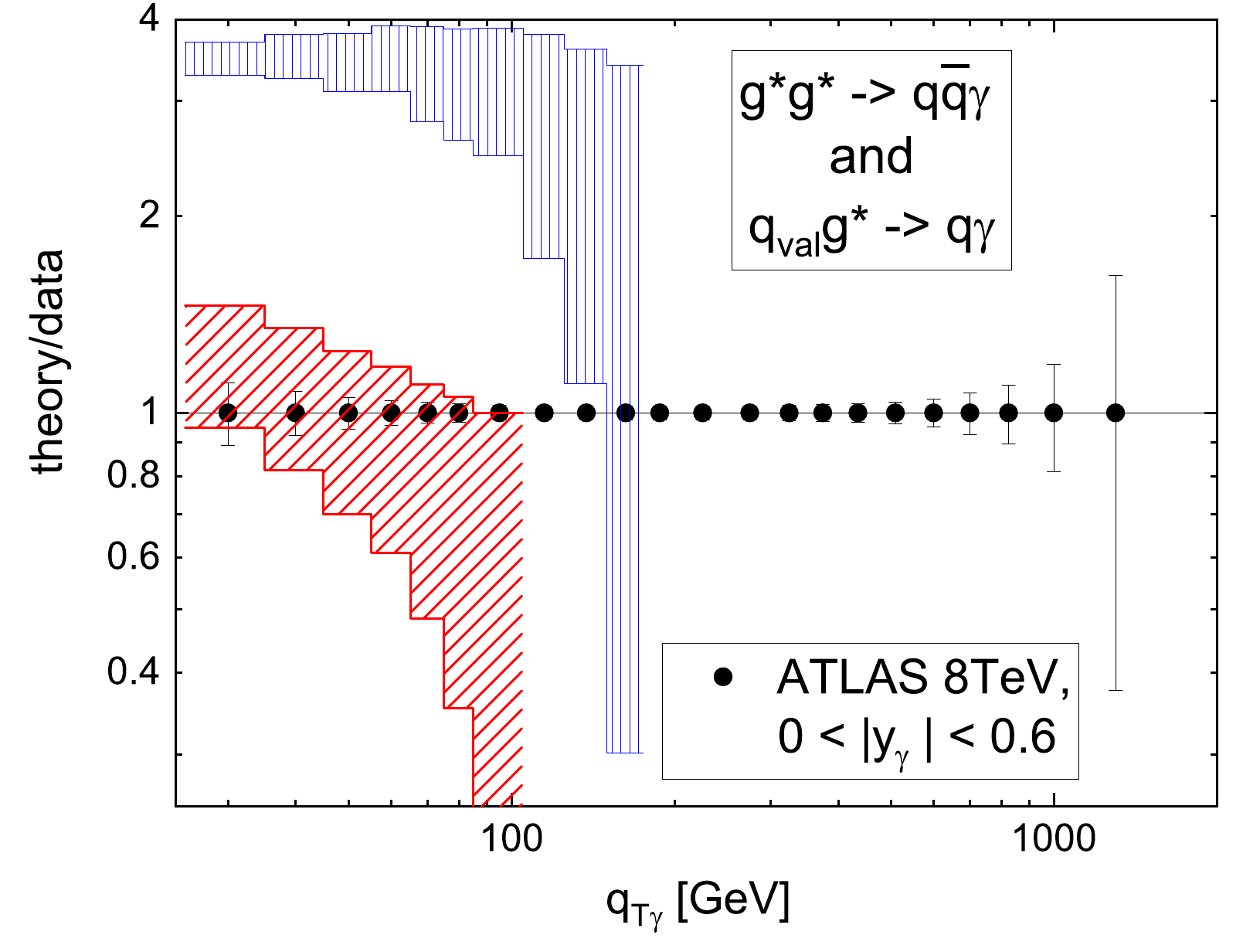}
\end{tabular}
\caption{
Theory to experiment ratios for the prompt photon production at ATLAS@8
TeV obtained from the JH and JH-2013 unintegrated gluon distributions
within the $qg^*$ (right) and $g^*g^*$ (left) schemes for the central rapidity
bin.
}
\label{fig:ratios_JH_JH2013}
\end{figure}

In Fig.\ \ref{fig:comp_ATLAS_8}, we show  the theory curves against
the ATLAS$@8{\rm TeV}$ data for the central rapidity, $|y|<0.6$, in the log-log scale. This is the way the experimental results are usually presented.
In the upper plot, the predictions are obtained in the $qg^*$ scheme while in the lower plot the predictions are computed in the $g^*g^*$ scheme. Clearly, the $qg^*$ scheme with the KMR-AO gluon distribution gives the best description of the data. The same gluon distribution used in the $g^*g^*$ scheme leads to a reasonable description of the data, but the cross section is somewhat underestimated for intermediate values of $q_T$. The cross sections obtained with the JH and KMR-AO distributions are very close to each other for $q_T < 100~{\rm GeV}$, where the JH parametrization may be applied.

The new JH-2013 parametrization allows to extend the JH results up to  $k_T \simeq 300~{\rm GeV}$.
However, the agreement with the data for $k_T< 100~{\rm GeV}$ becomes much worse in the $g^*g^*$ scheme.
 The predictions obtained with the GBW gluon are close to the data at lower $q_T$, but lead to too flat $q_T$-dependence. Clearly, this is due to too flat $x$-dependence of the GBW gluon, discussed in more detail in the previous section.  Based on Fig.~\ref{fig:comp_ATLAS_8}, one concludes that the best description of the prompt photon data from the LHC
 in the full range  of the measured photon transverse momentum  is obtained   for the KMR-AO gluon  distribution in both the  $qg^*$ and $g^*g^*$ schemes.

\begin{figure}[t]
\begin{tabular}{lll}
\hskip -4mm
\includegraphics[width=0.35\textwidth]{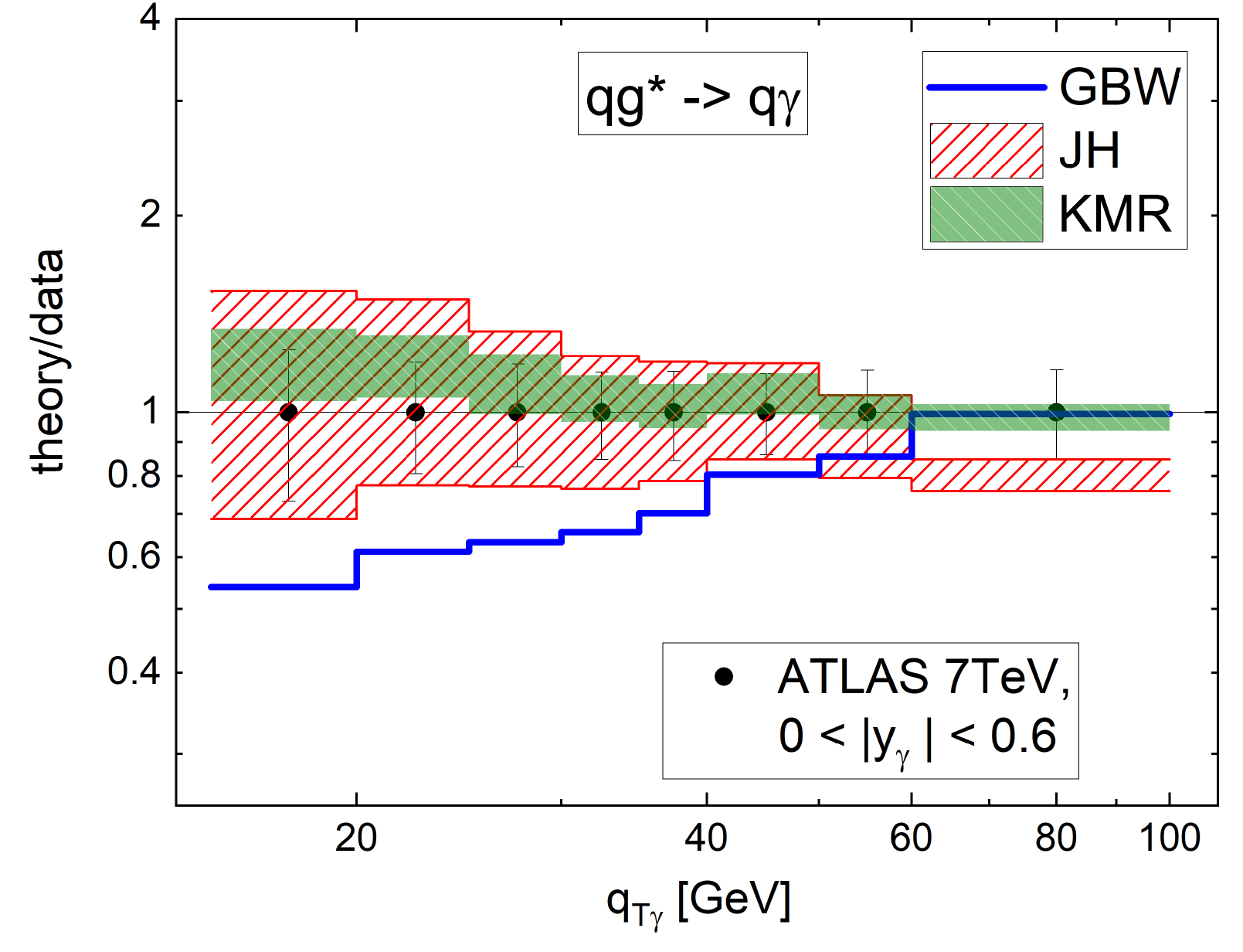}  &
\hskip -3mm
\includegraphics[width=0.35\textwidth]{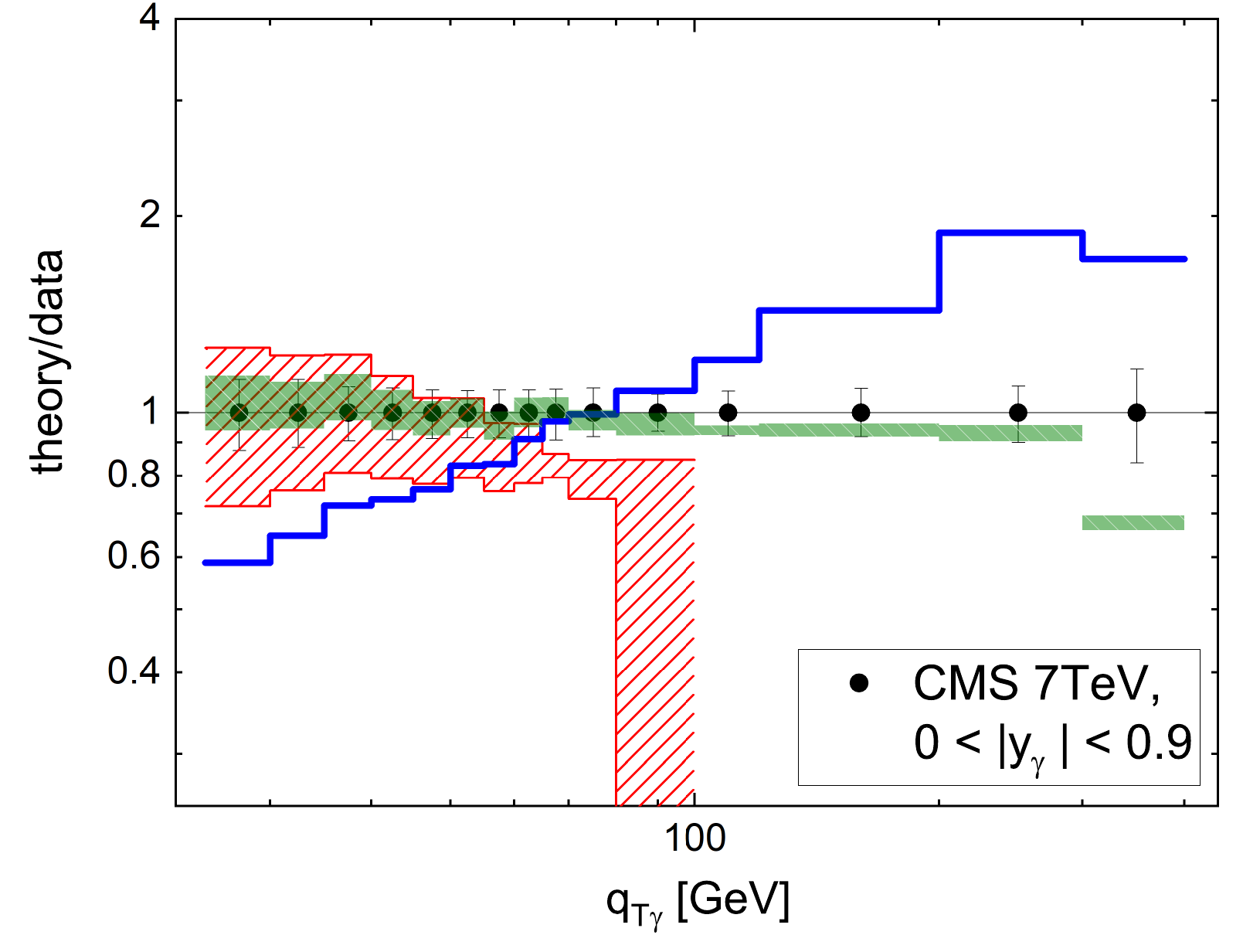}    &
\hskip -3mm
\includegraphics[width=0.35\textwidth]{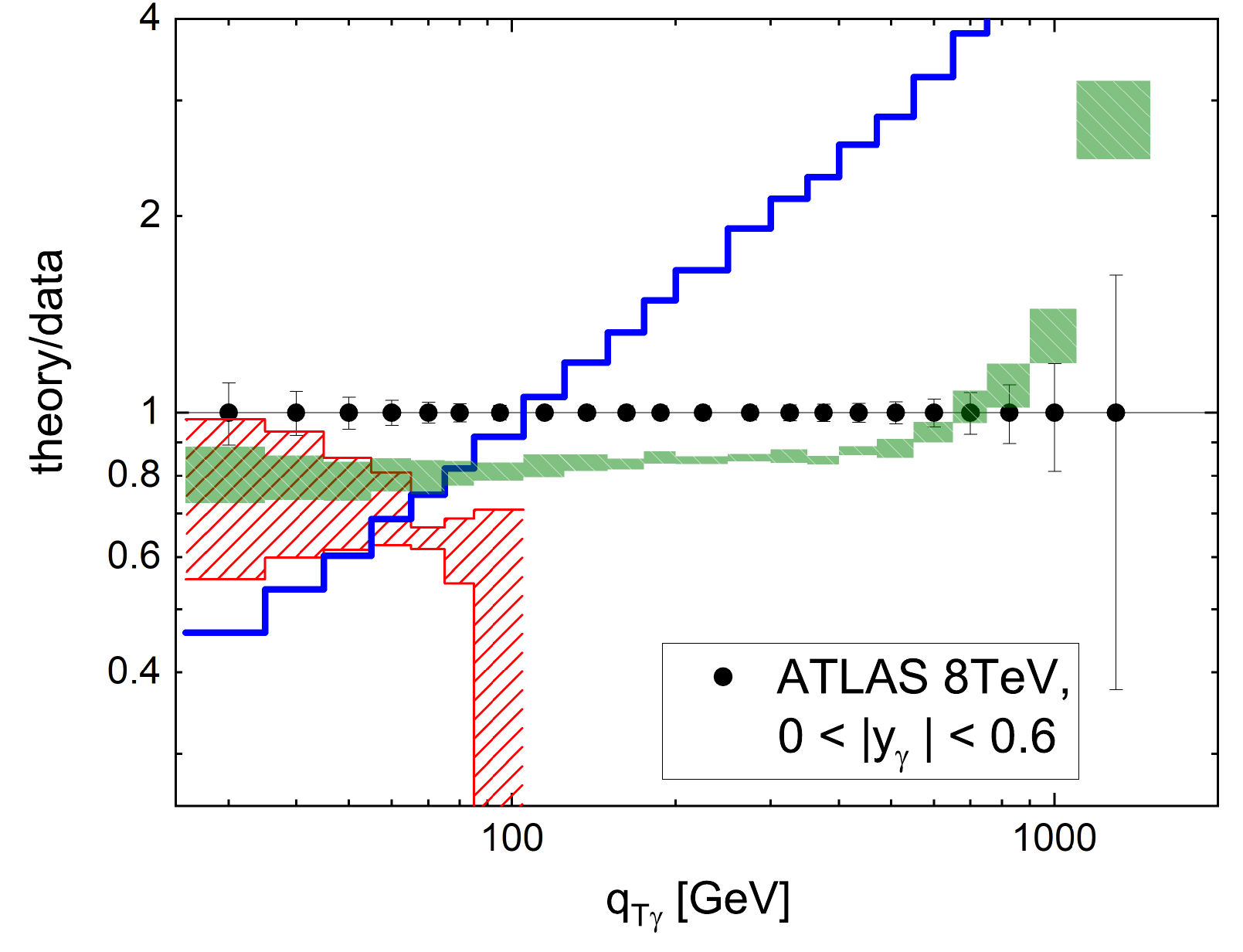}\\
\hskip -4mm
\includegraphics[width=0.35\textwidth]{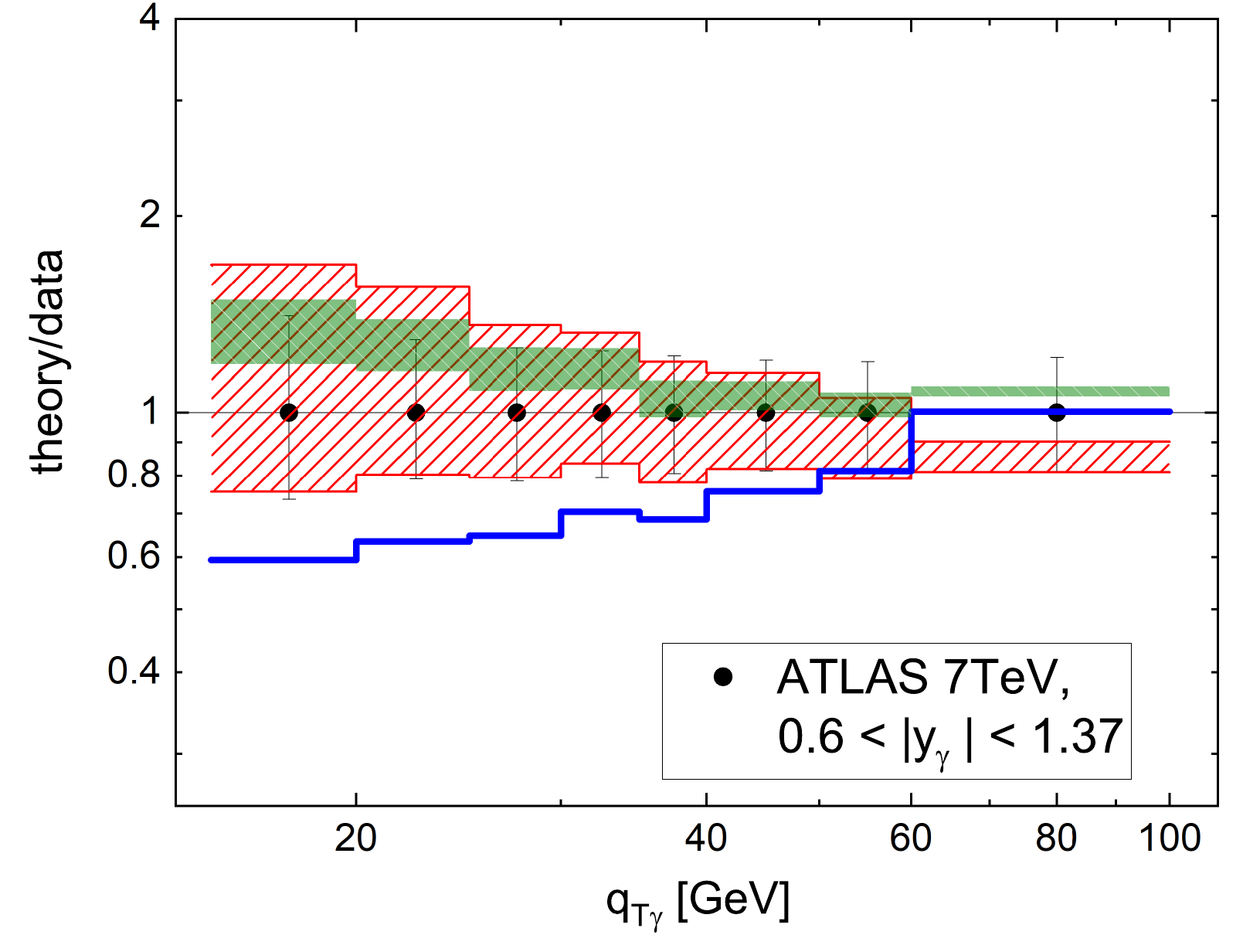} & 
\hskip -3mm
\includegraphics[width=0.35\textwidth]{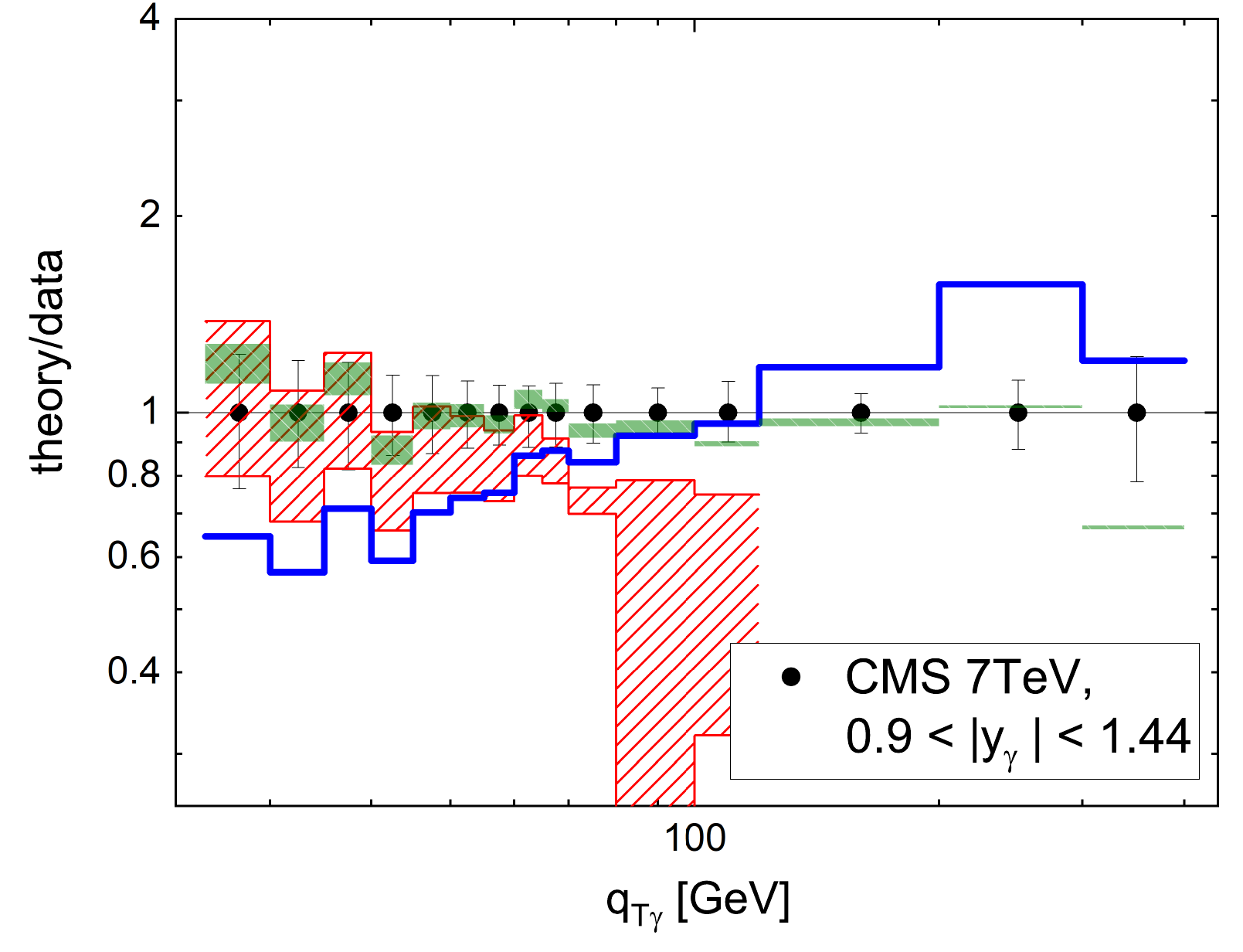}   &
\hskip -3mm
\includegraphics[width=0.35\textwidth]{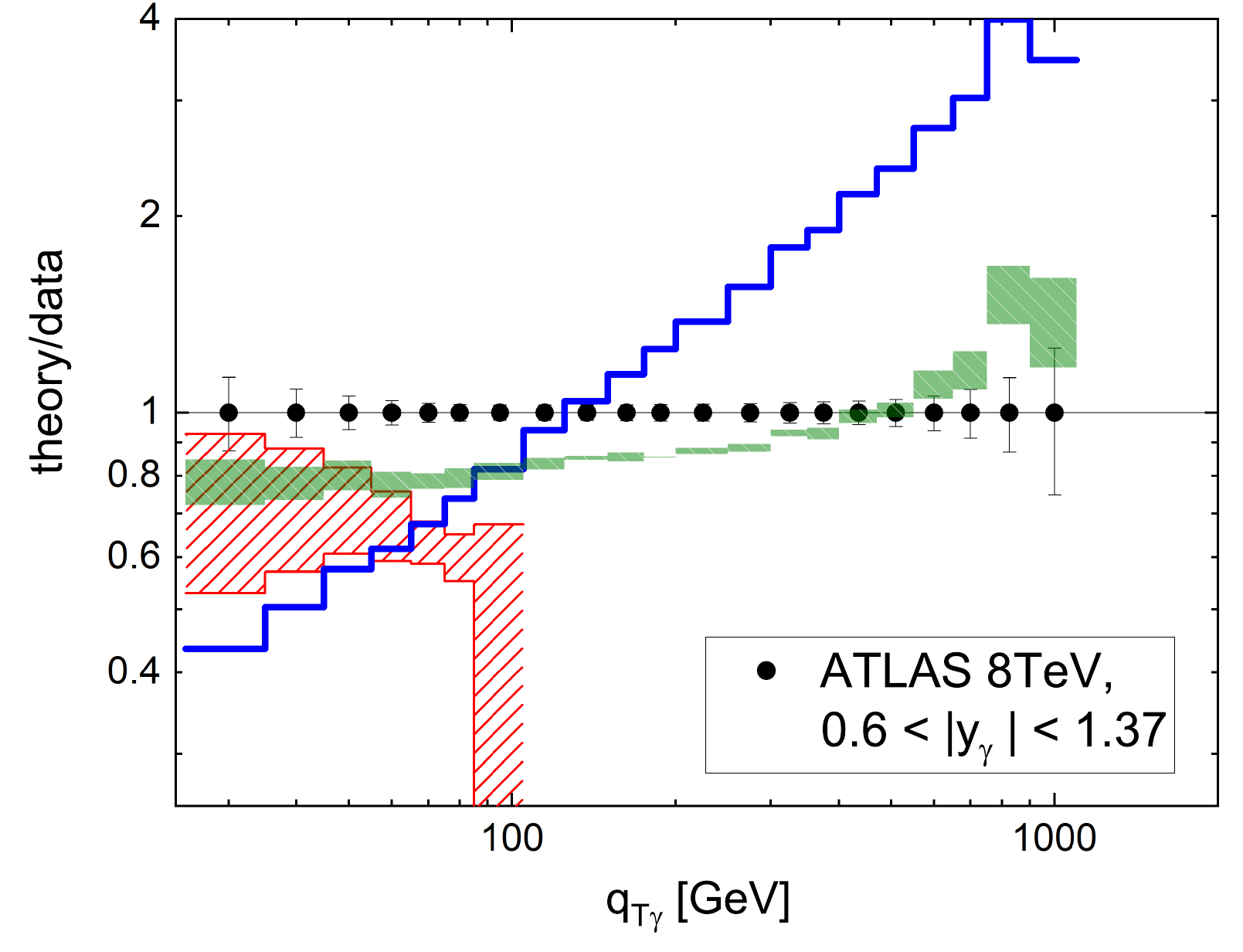}\\
\hskip -4mm
\includegraphics[width=0.35\textwidth]{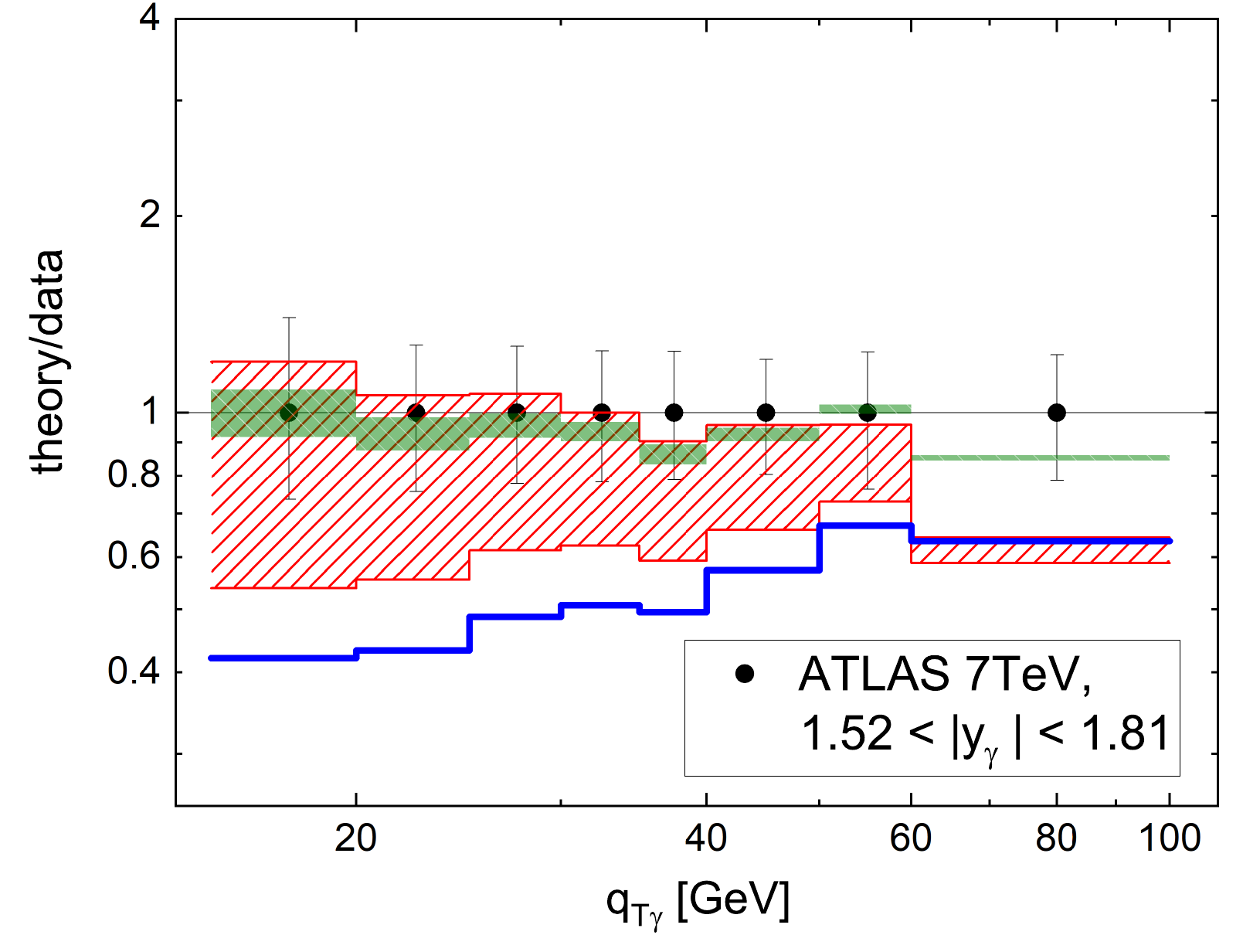} &
\hskip -3mm
\includegraphics[width=0.35\textwidth]{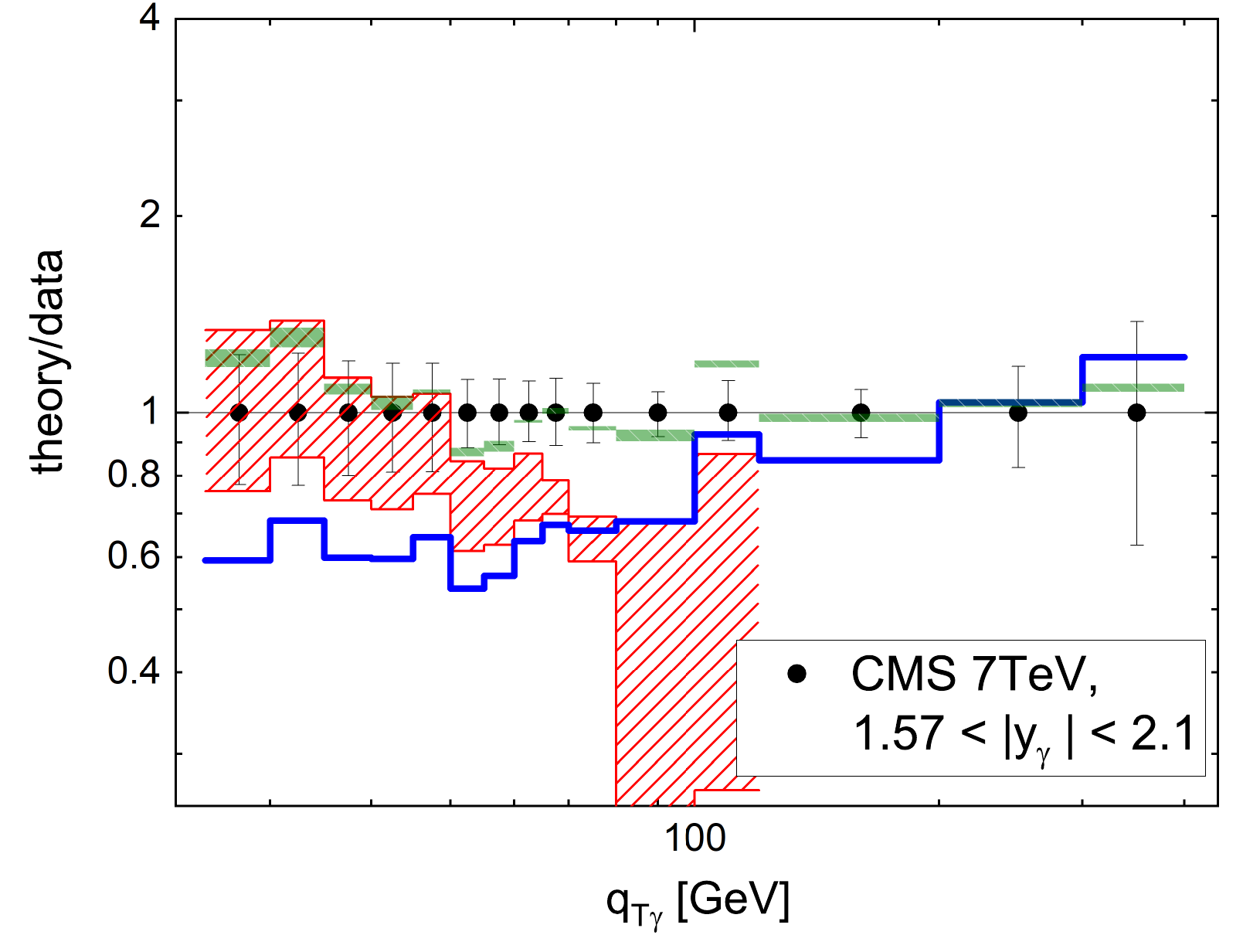}   &
\hskip -3mm
\includegraphics[width=0.35\textwidth]{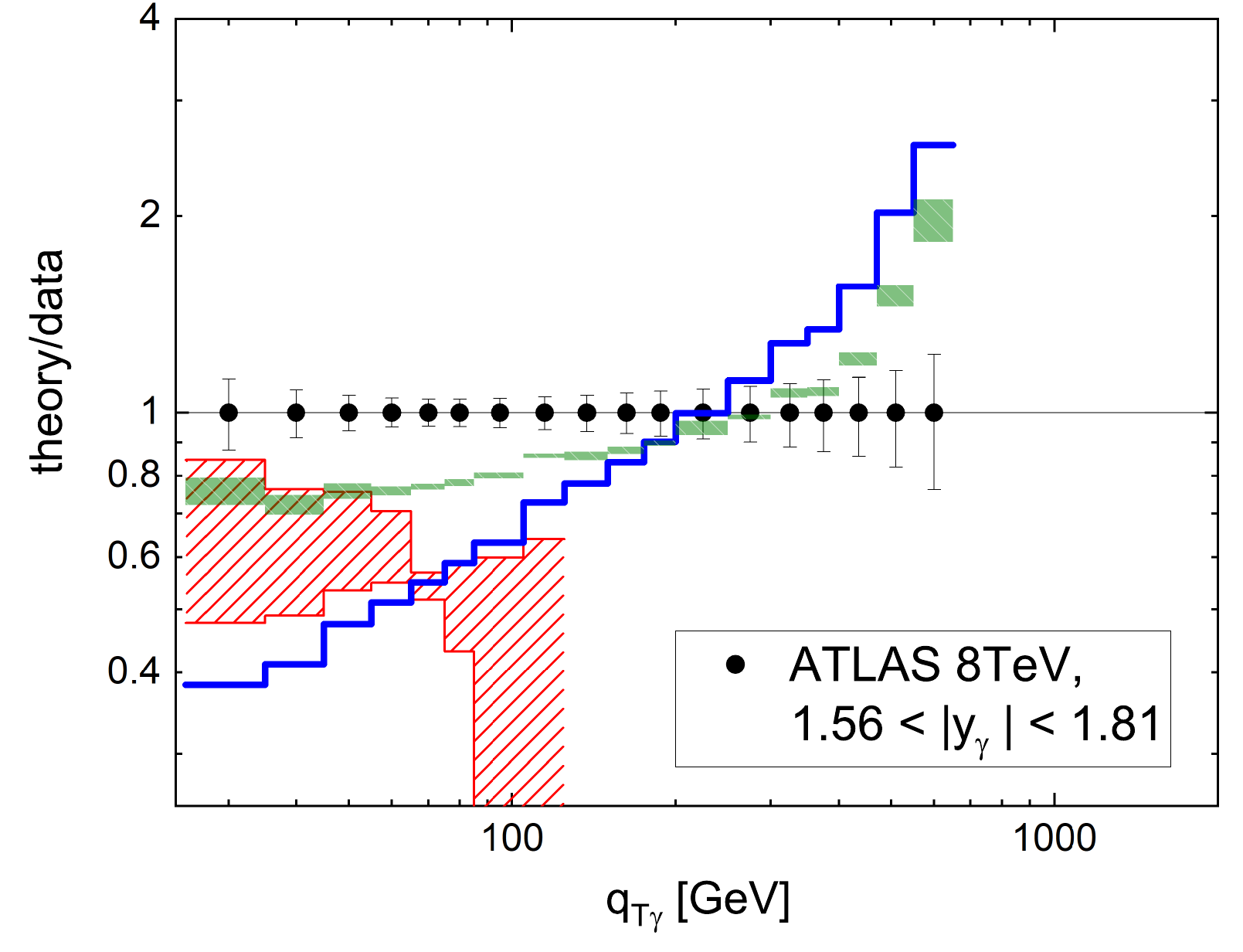}\\
&
\hskip -3mm
\includegraphics[width=0.35\textwidth]{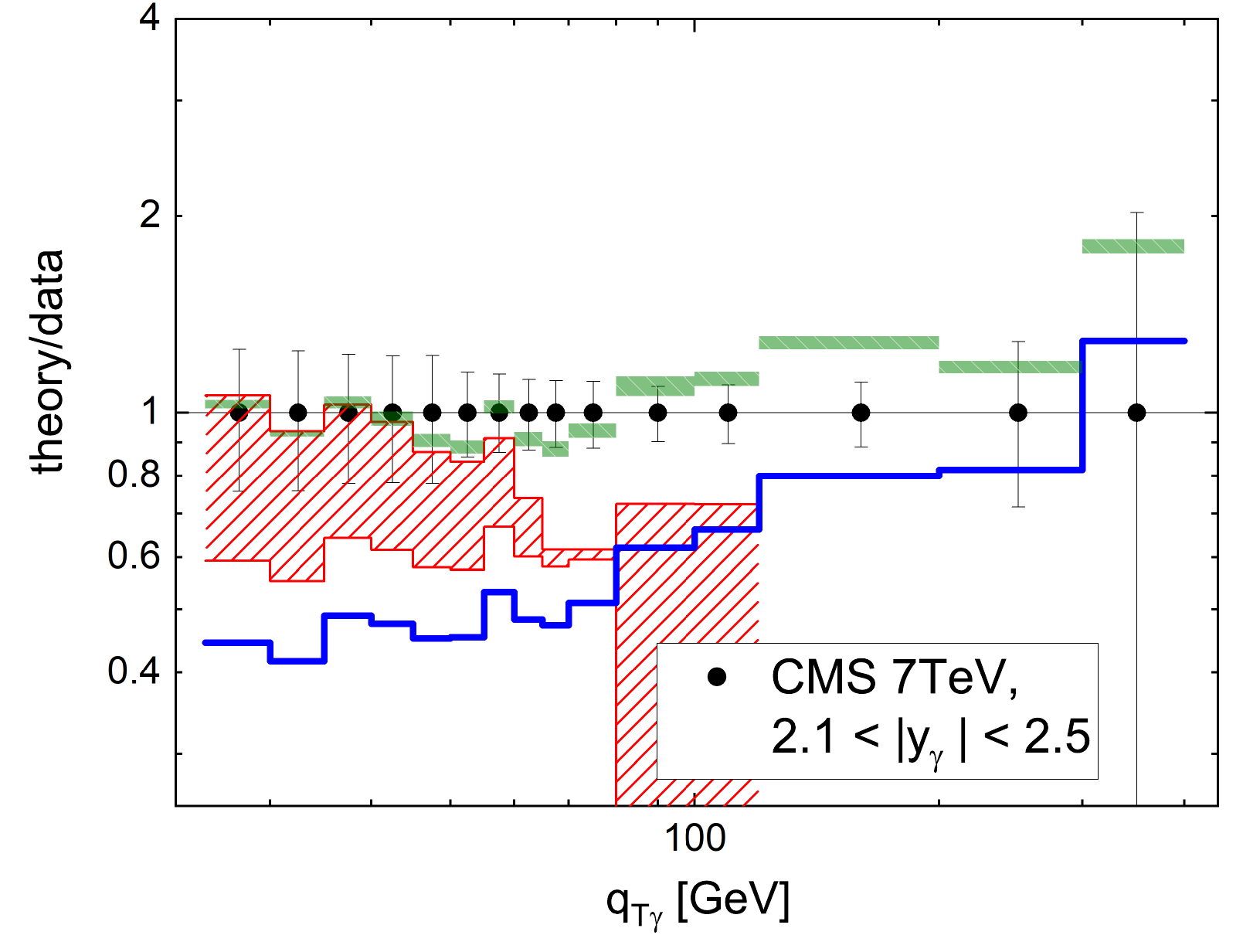} &
\hskip -3mm
\includegraphics[width=0.35\textwidth]{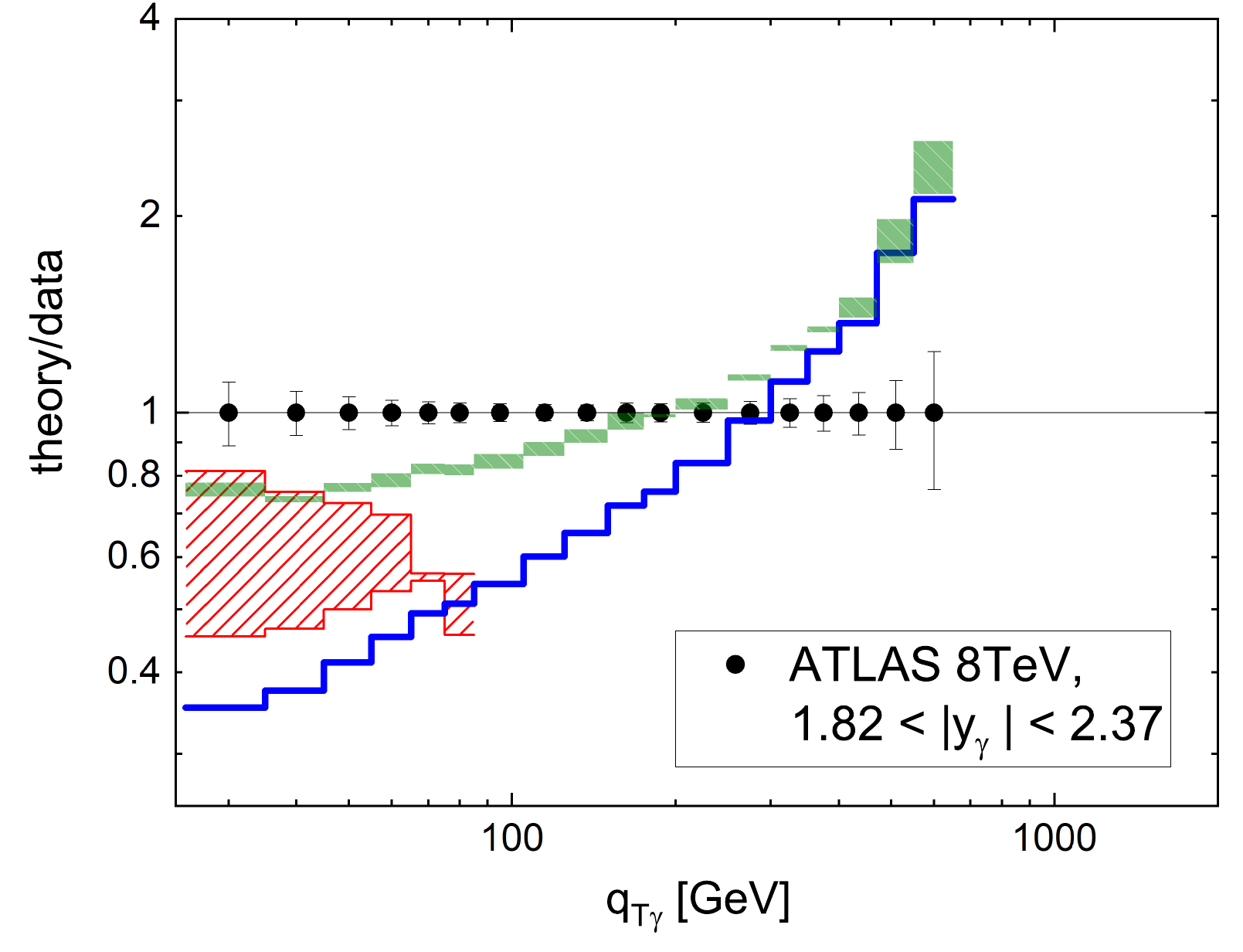} \\
\end{tabular}
\caption{Theory to experiment ratios for the prompt photon production  within the $qg^*$ scheme for  the
ATLAS$@7{\rm TeV}$ \cite{Aad:2010sp} (left column), CMS$@7{\rm TeV}$ \cite{Chatrchyan:2011ue} (central column) and ATLAS$@8{\rm TeV}$ \cite{Aad:2016xcr} (right column) data and 
the KMR-AO (green shaded bands), JH (red hatched bands) and GBW (blue curves) unintegrated gluon distributions. The theory uncertainties are due to
the variation of the factorization  scale $\mu_F=\mu_R$ between $q_T/2$ and $2q_T$.}
\label{fig:comp_ATLAS_CMS_78ratio_qg}
\end{figure}

\begin{figure}[t]
\begin{tabular}{lll}
\hskip -4mm
\includegraphics[width=0.35\textwidth]{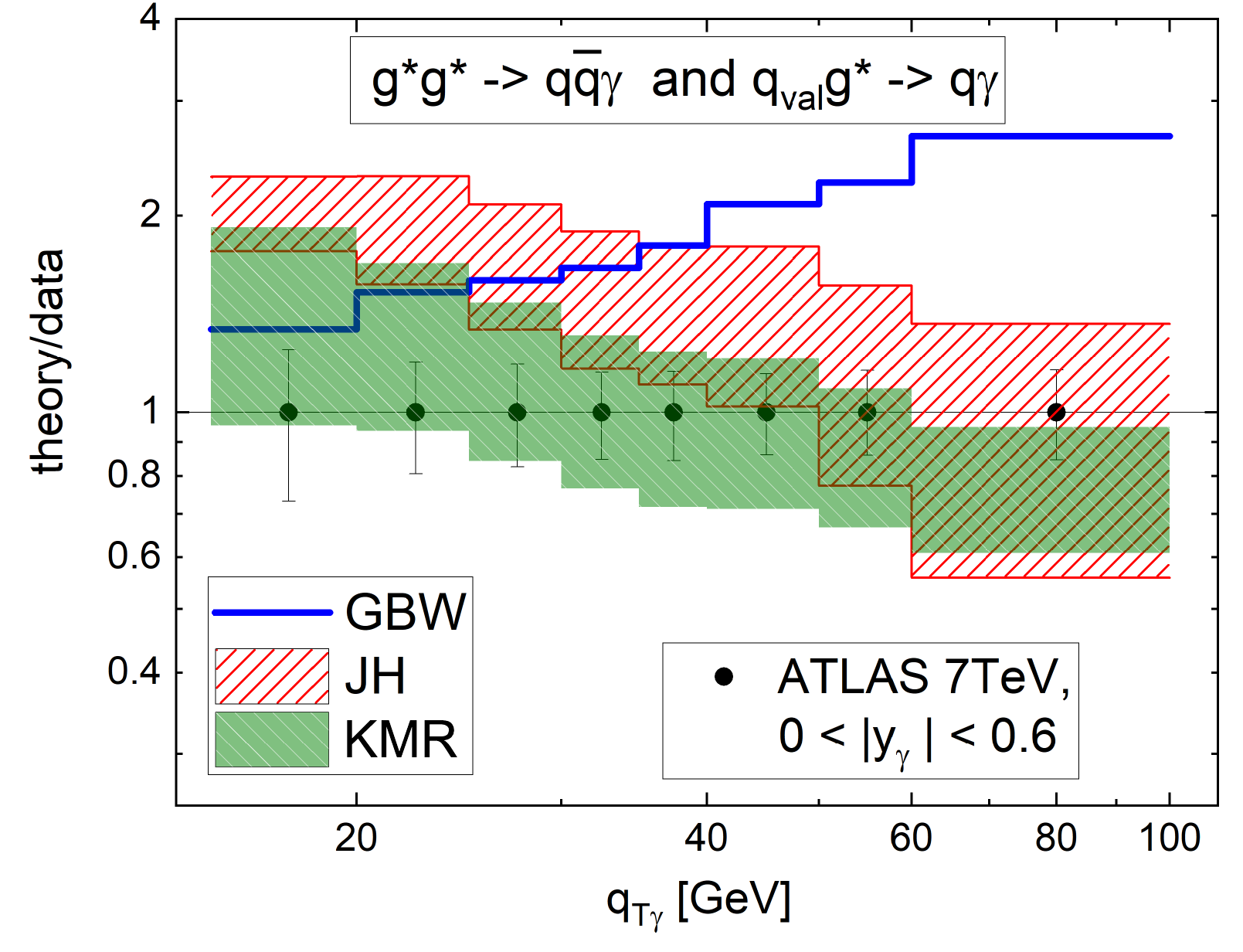} &
\hskip -3mm
\includegraphics[width=0.35\textwidth]{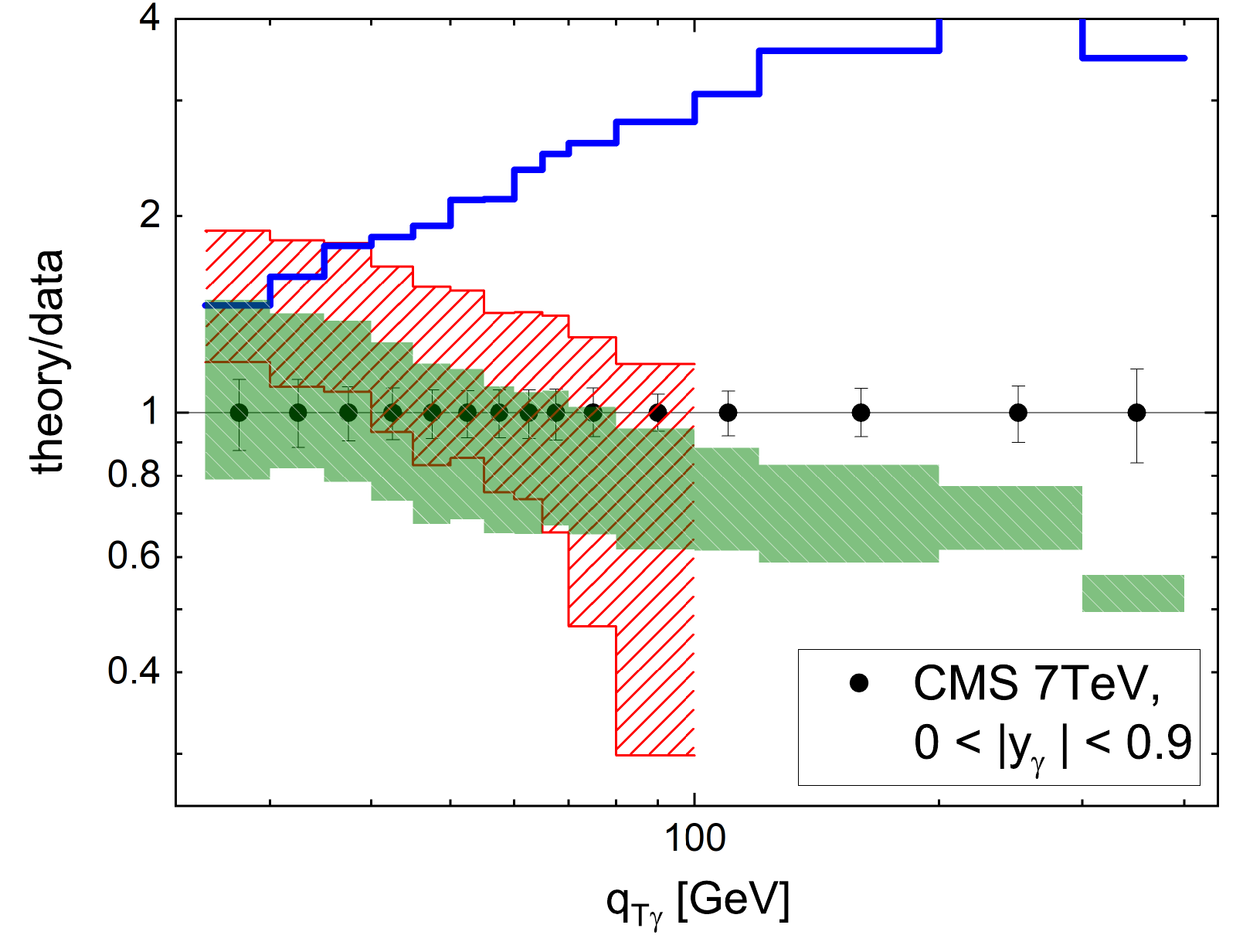}    &
\hskip -3mm
\includegraphics[width=0.35\textwidth]{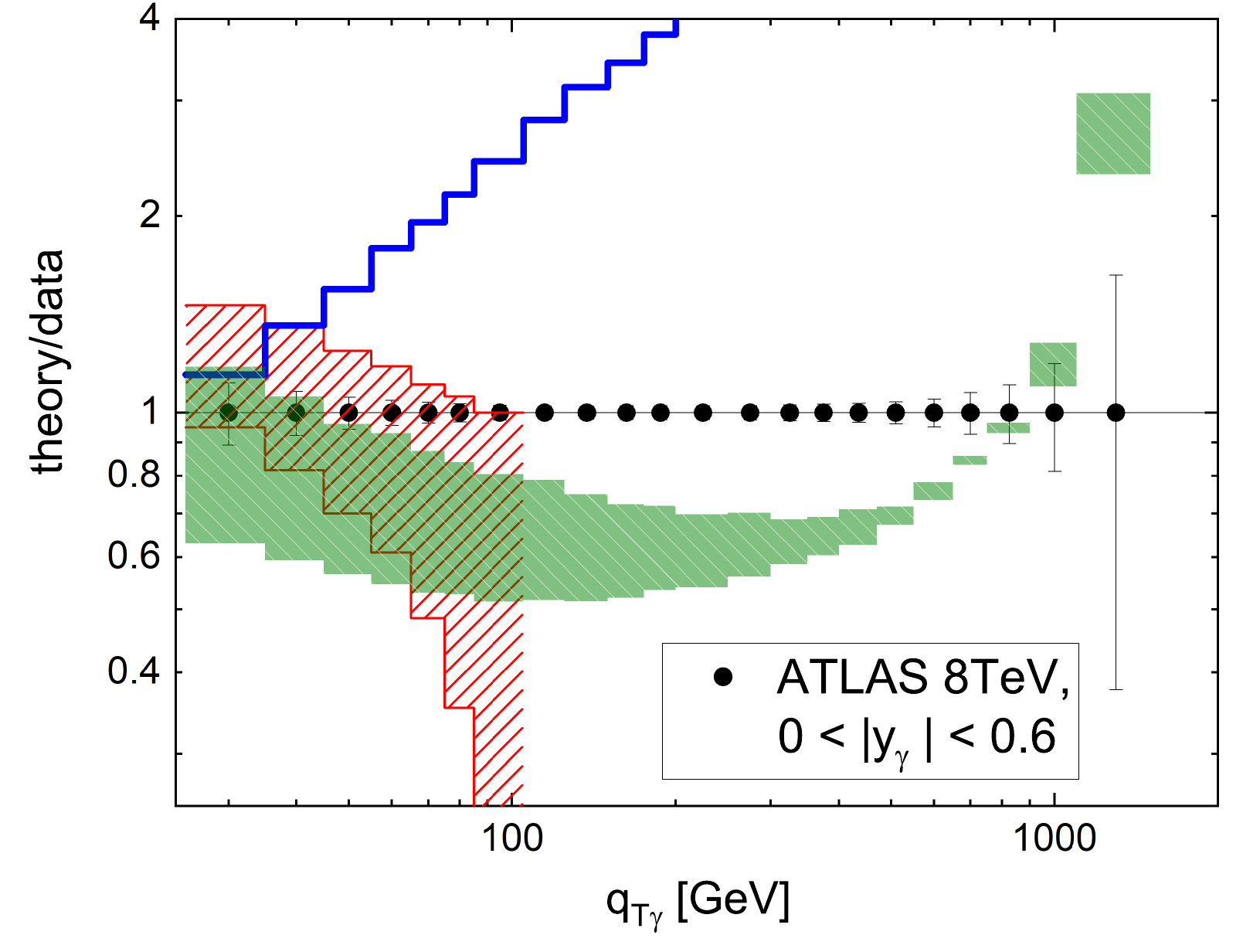} \\
\hskip -4mm
\includegraphics[width=0.35\textwidth]{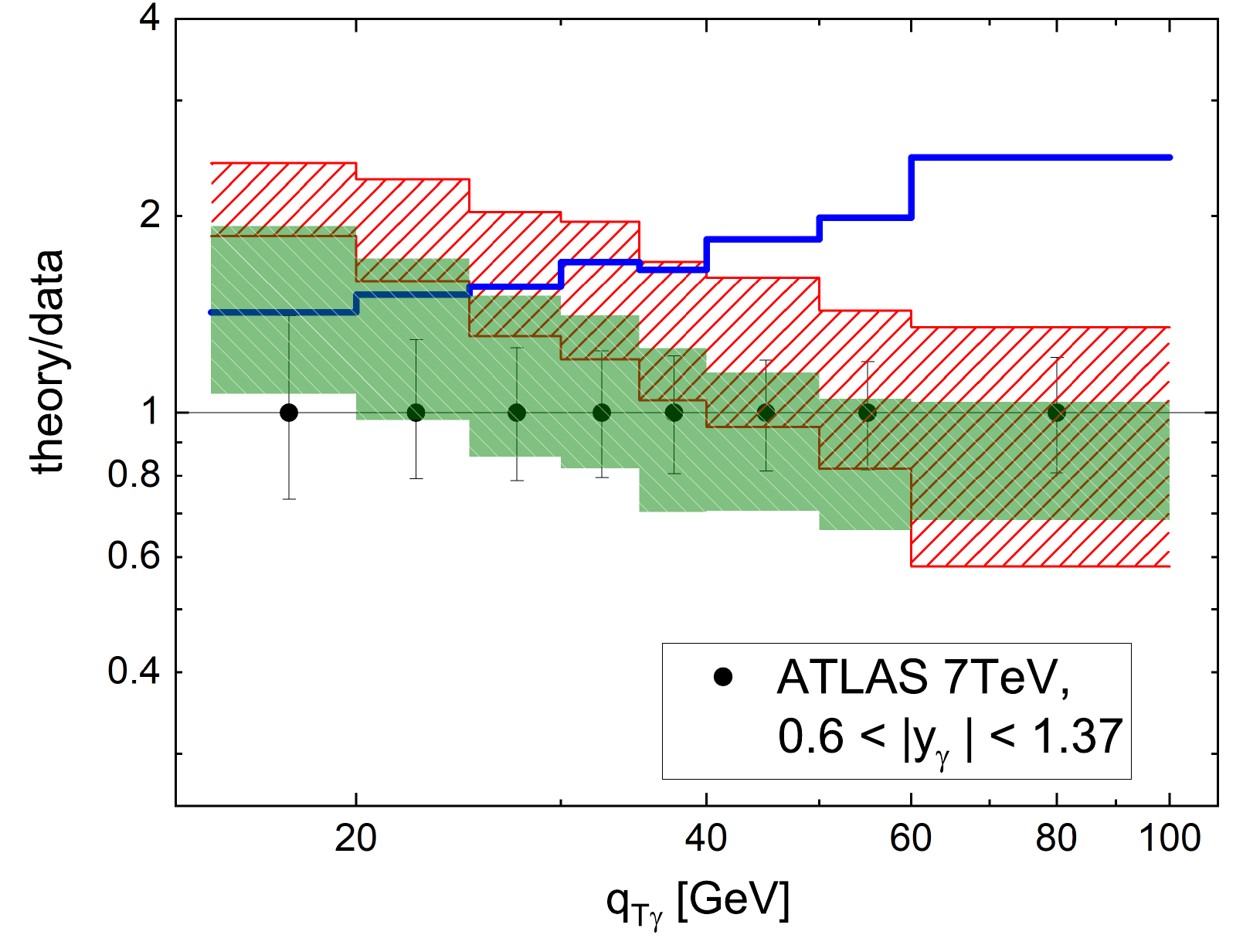} &
\hskip -3mm
\includegraphics[width=0.35\textwidth]{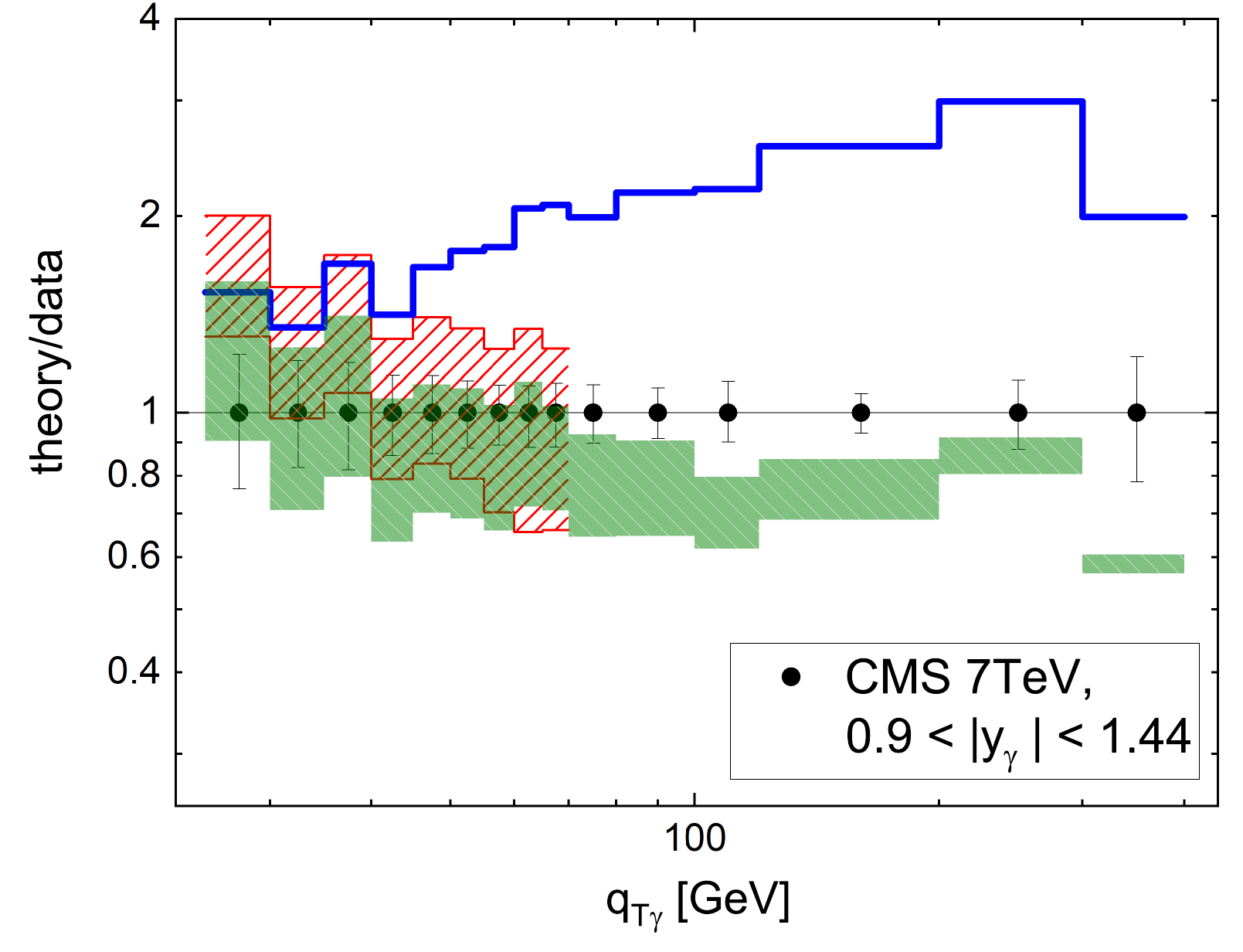}   &
\hskip -3mm
 \includegraphics[width=0.35\textwidth]{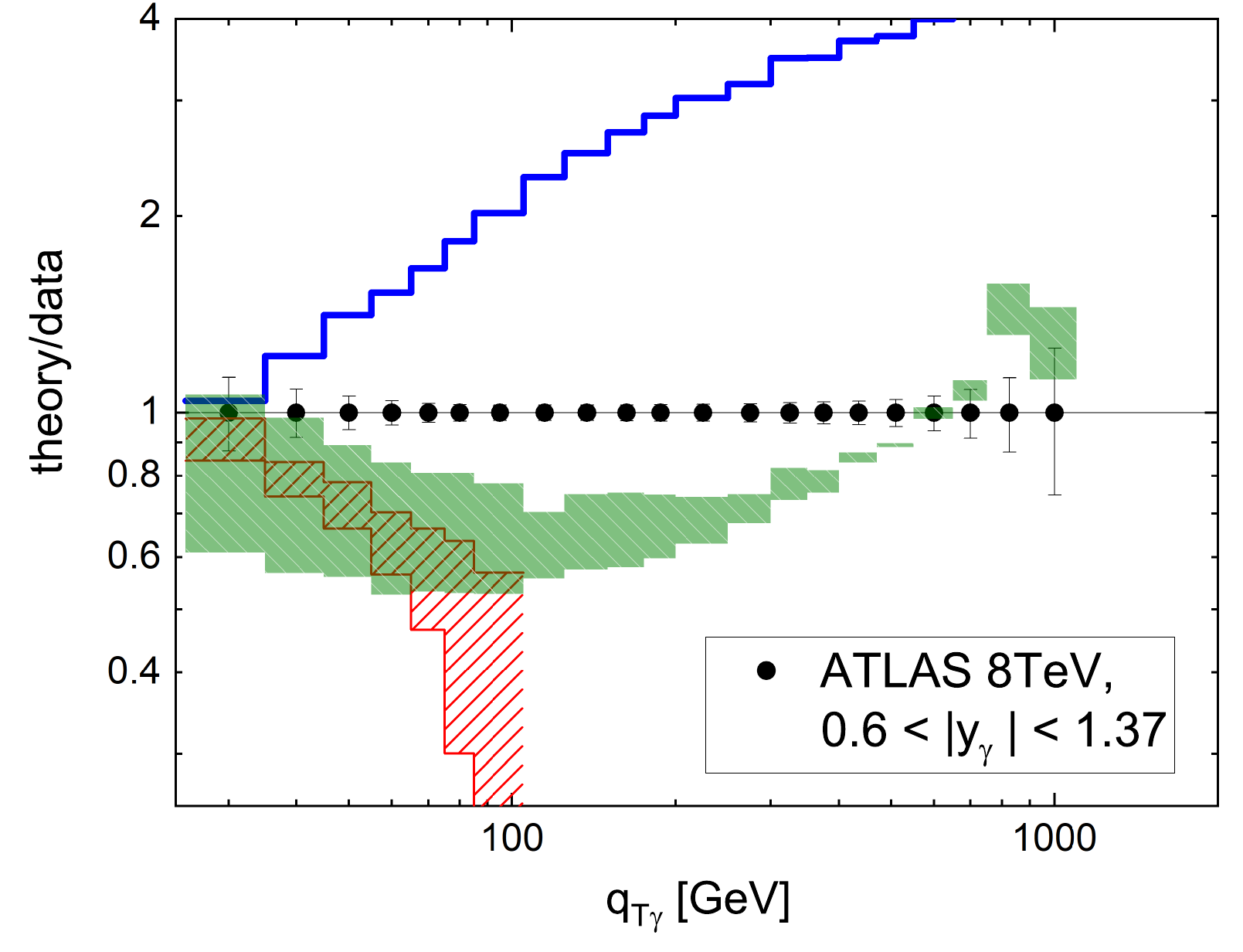} \\
\hskip -4mm
\includegraphics[width=0.35\textwidth]{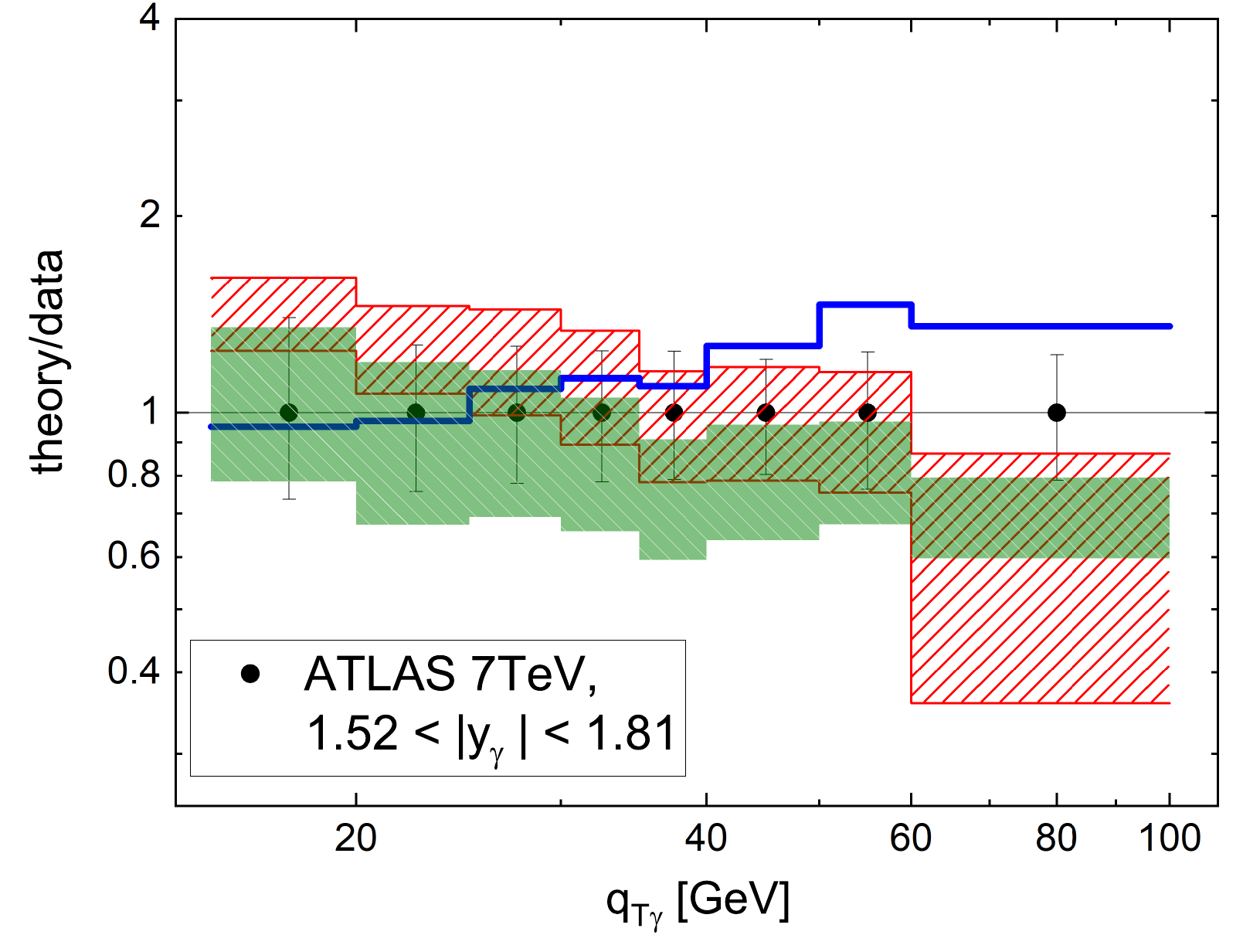} &
\hskip -3mm
\includegraphics[width=0.35\textwidth]{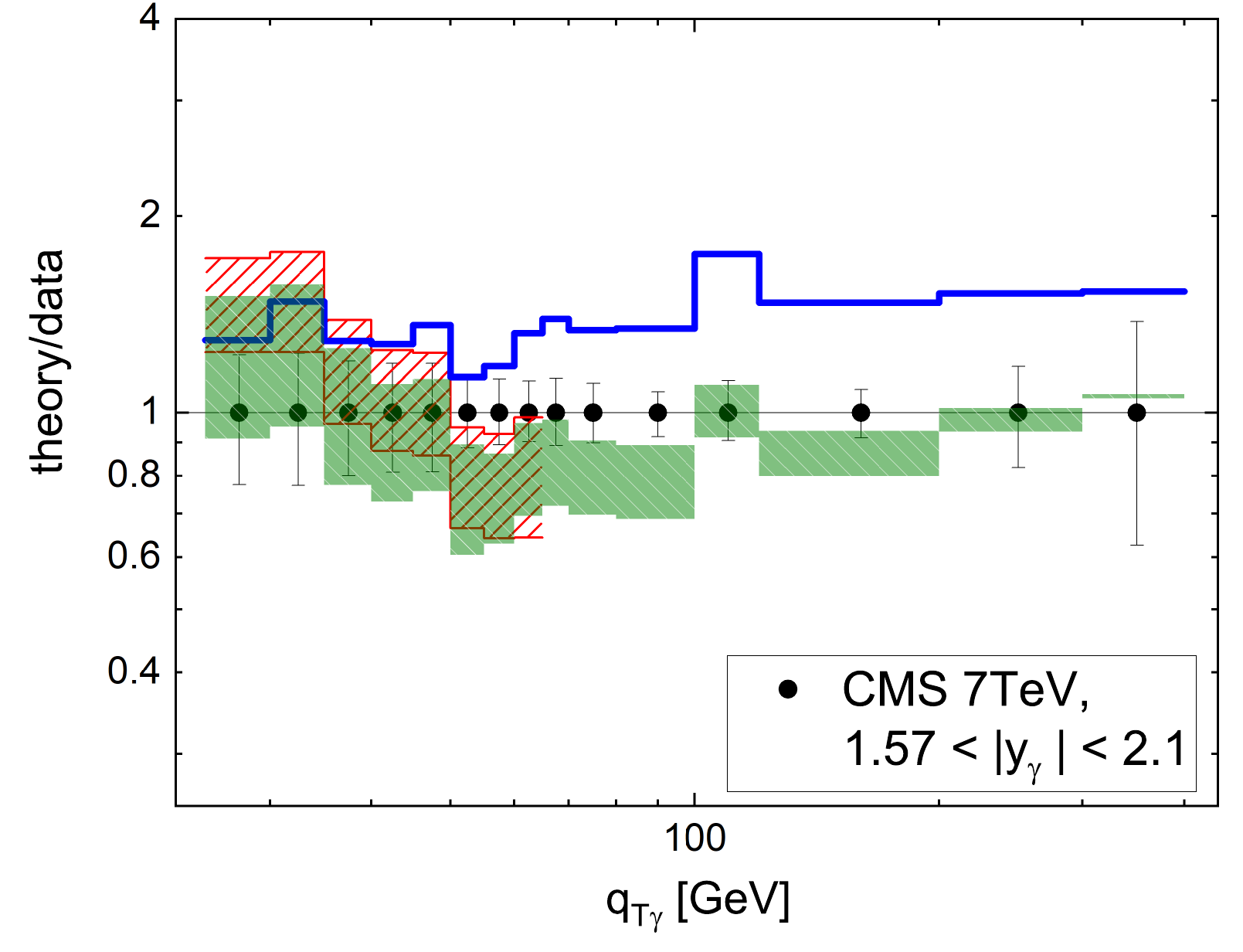}   &
\hskip -3mm
\includegraphics[width=0.35\textwidth]{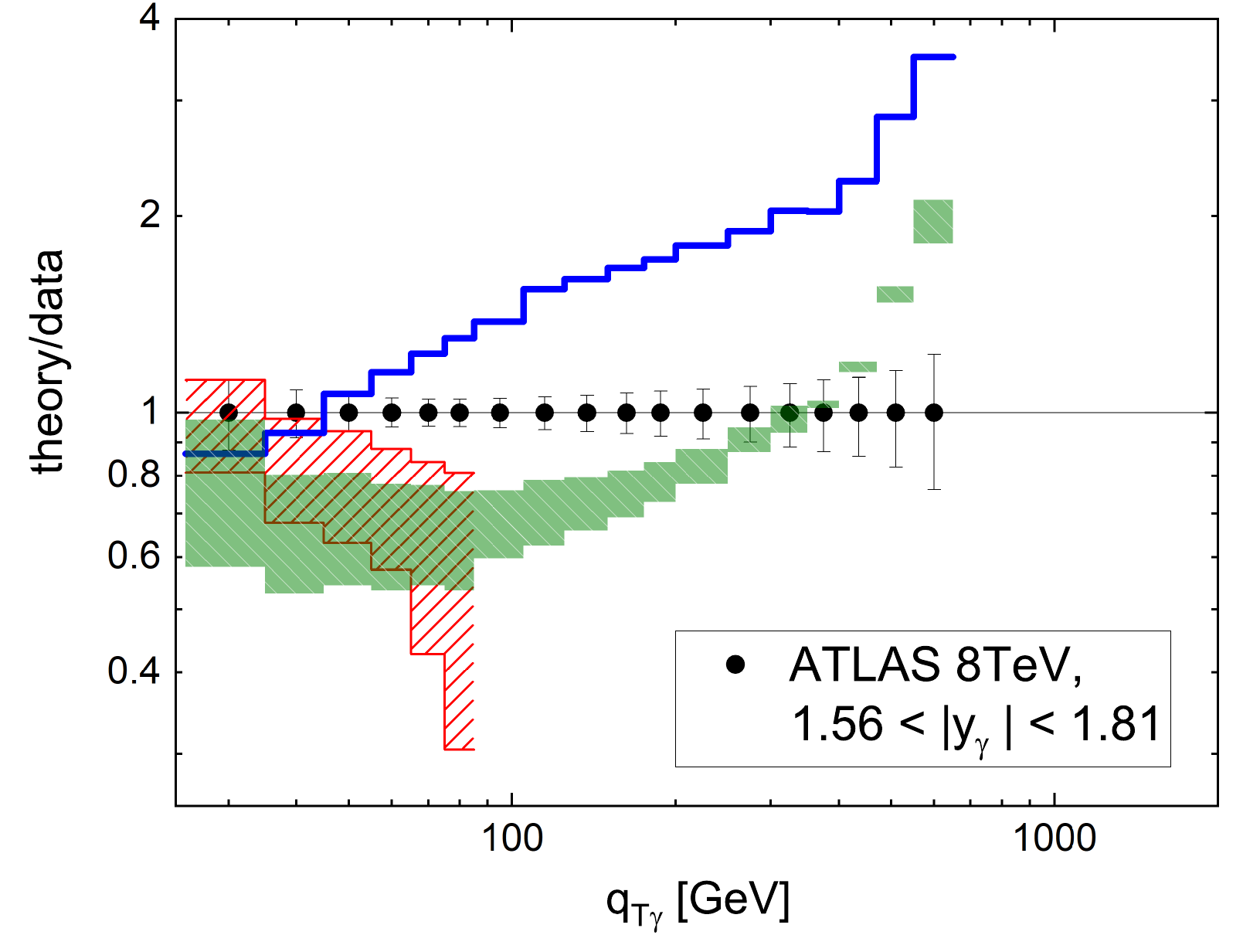} \\
&
\hskip -3mm
\includegraphics[width=0.35\textwidth]{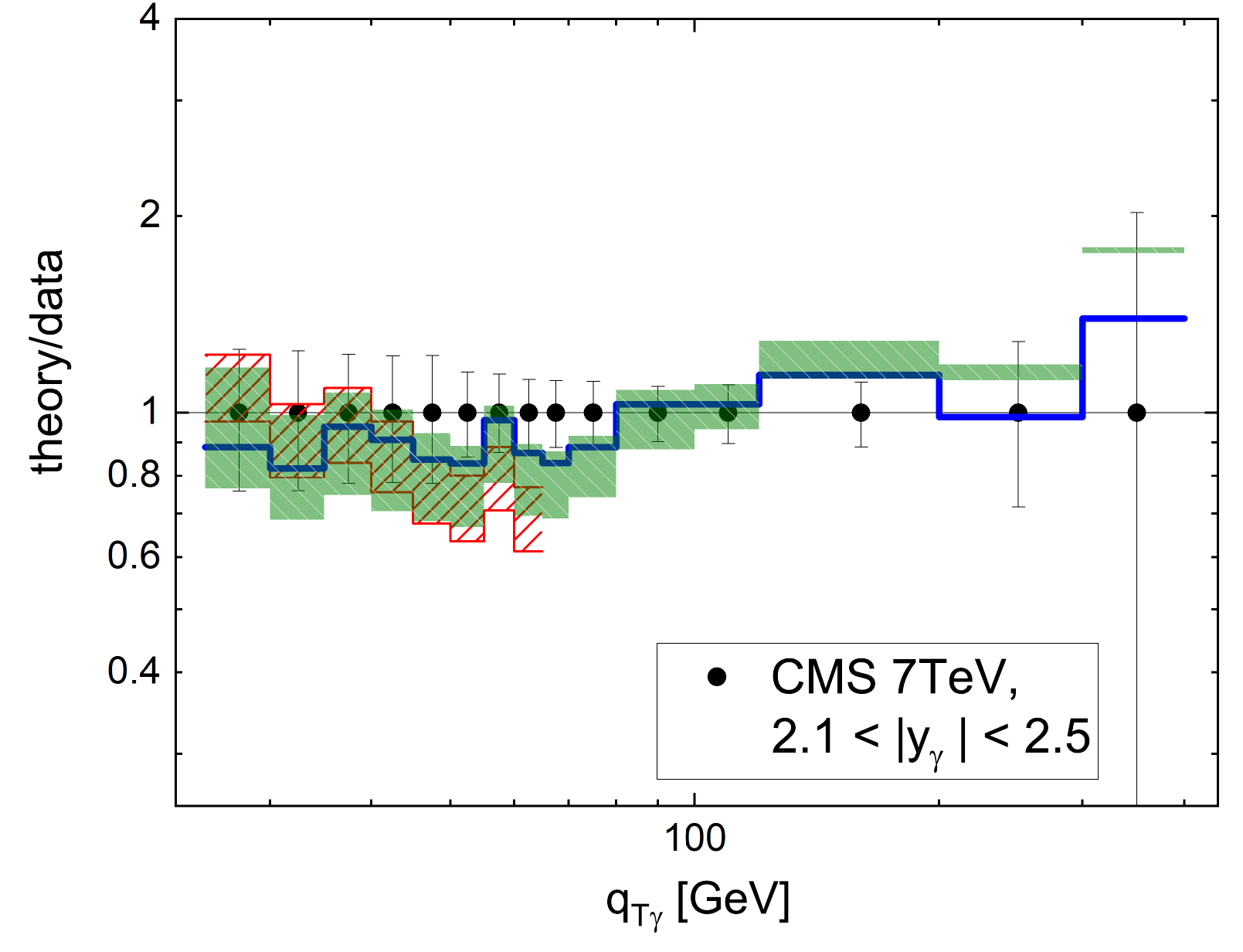}  &
\hskip -3mm
\includegraphics[width=0.35\textwidth]{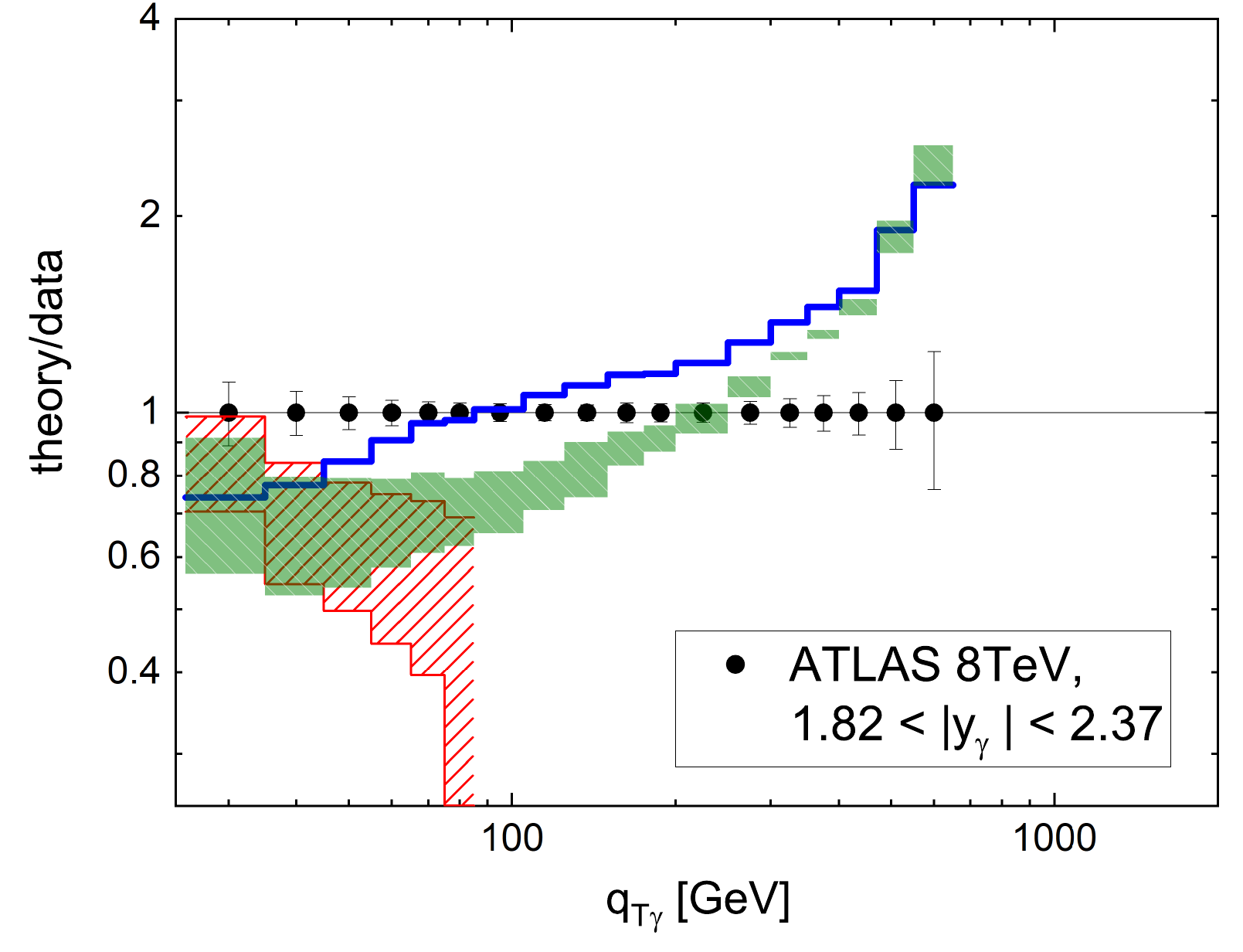}\\
\end{tabular}
\caption{Theory to experiment ratios for the prompt photon production  within the $g^*g^*$ scheme for  the
ATLAS$@7{\rm TeV}$ \cite{Aad:2010sp} (left column), CMS$@7{\rm TeV}$ \cite{Chatrchyan:2011ue} (central column) and ATLAS$@8{\rm TeV}$ \cite{Aad:2016xcr} (right column) data and 
the KMR-AO (green shaded  bands), JH (red hatched bands) and GBW (blue curves) unintegrated gluon distributions. The theory uncertainties
are due to the variation of the factorization  scale $\mu_F=\mu_R$ between $q_T/2$ and $2q_T$.}
\label{fig:comp_ATLAS_CMS_78ratio_gg}
\end{figure}

In order to perform more detailed studies of the description quality, we change the way the data and theoretical results are presented. Namely, in Figs.~\ref{fig:ratios_JH_JH2013},~\ref{fig:comp_ATLAS_CMS_78ratio_qg} and \ref{fig:comp_ATLAS_CMS_78ratio_gg} we display the  ratios of 
the theoretical results to the experimental data. For the  sake of the optimal comparison, the theoretical result are integrated over the bins used in the experiments.  The theoretical uncertainties are due to the variation of the  factorization scales, 
$\mu_F=\mu_R$, between $q_T/2$ and $2q_T$.

 In Fig.\ \ref{fig:ratios_JH_JH2013}, using central bin of the ATLAS$@8{\rm TeV}$ data, we show the typical result for the comparison of the data with the predictions given by the JH and JH-2013 gluon distributions. We see that the new JH-2013 distribution significantly overestimates the data for $q_T< 100~{\rm GeV}$.  This could be attributed to a much higher the JH-2013 gluon distribution than the JH one for the gluon $k_T<1~{\rm GeV}$, see Fig.\ \ref{updf_comparison}. In the forthcoming
 analyses we will show only the JH predictions.

The theory to data ratios obtained within the $qg^*$ scheme are shown in Fig.~\ref{fig:comp_ATLAS_CMS_78ratio_qg} for the three considered
LHC data sets. The absolute value of the photon rapidities $|y|$ grows from the top row down.
Clearly, the best overall description of the data is found with the KMR-AO gluon (green shaded bands). It is seen at best for  the data set with the smallest errors, that is the ATLAS$@8{\rm TeV}$ data (right column). For most bins, these theoretical results are close to the data, with
relative differences smaller than 20\%.  Given the fact that the leading order $k_T$-factorization approach  is used, this is a very good accuracy. 
Somewhat larger deviations are found only for larger values of $|y|$ and $q_T>500$~GeV.  In this region, however, the $k_T$-factorization approach is less accurate since the contributions from relatively large values of the gluon $x$ are becoming important. The scale uncertainty of these results is very small and does not allow to overlap with the data.  The description of the CMS$@7{\rm TeV}$ data with $q_T<400$~GeV (middle column)  with the KMR-AO gluon  is  also very good for all rapidities. The predictions for ATLAS$@7{\rm TeV}$ (left column) are above the central data points, 
but stays consistent with the data within the experimental errors, somewhat larger in this measurement. 

The data description in Fig.~\ref{fig:comp_ATLAS_CMS_78ratio_qg} with the JH gluon (red hatched bands) is reasonable, but the application of this gluon is limited to $q_T <100$~GeV. The results calculated with JH are somewhat below the KMR-AO results and more away from the  CMS$@7{\rm TeV}$ and ATLAS$@8{\rm TeV}$ data. This is interesting in the context of Fig.~\ref{pdf_comparison}, where one sees a larger integrated gluon from JH than from KMR-AO distributions. This suggests that the lower prompt photon production cross sections from the JH distribution are due to the cut-off of larger transverse momenta, $k_T>\mu_F$, that is not present in the KMR-AO case, see Fig.~\ref{updf_comparison}. This shows that the prompt photon data are sensitive to the shape of the transverse momentum gluon distribution, and that the longer  $k_T$-tail of the KMR-AO distribution is preferred by the data.
 
We also see that the uncertainty of the cross section due to scale variation is much larger for JH than for KMR-AO. The gluon from the GBW model does not reproduce well the $q_T$-dependence of the data. The predictions are significantly below the data at the lower limits of $q_T$ and much above data at very large~$q_T$. One has to remember, however, that in our study the GBW model is used far outside the range of very small~$x$ values and moderate scales, where it was fitted and should work well. The deviations from the data reflect the tendency visible already in the integrated GBW gluon, see Fig.~\ref{pdf_comparison} and the corresponding discussion of its content.

\begin{figure}
\begin{tabular}{ccc}
\hskip -3mm
\includegraphics[width=0.35\textwidth]{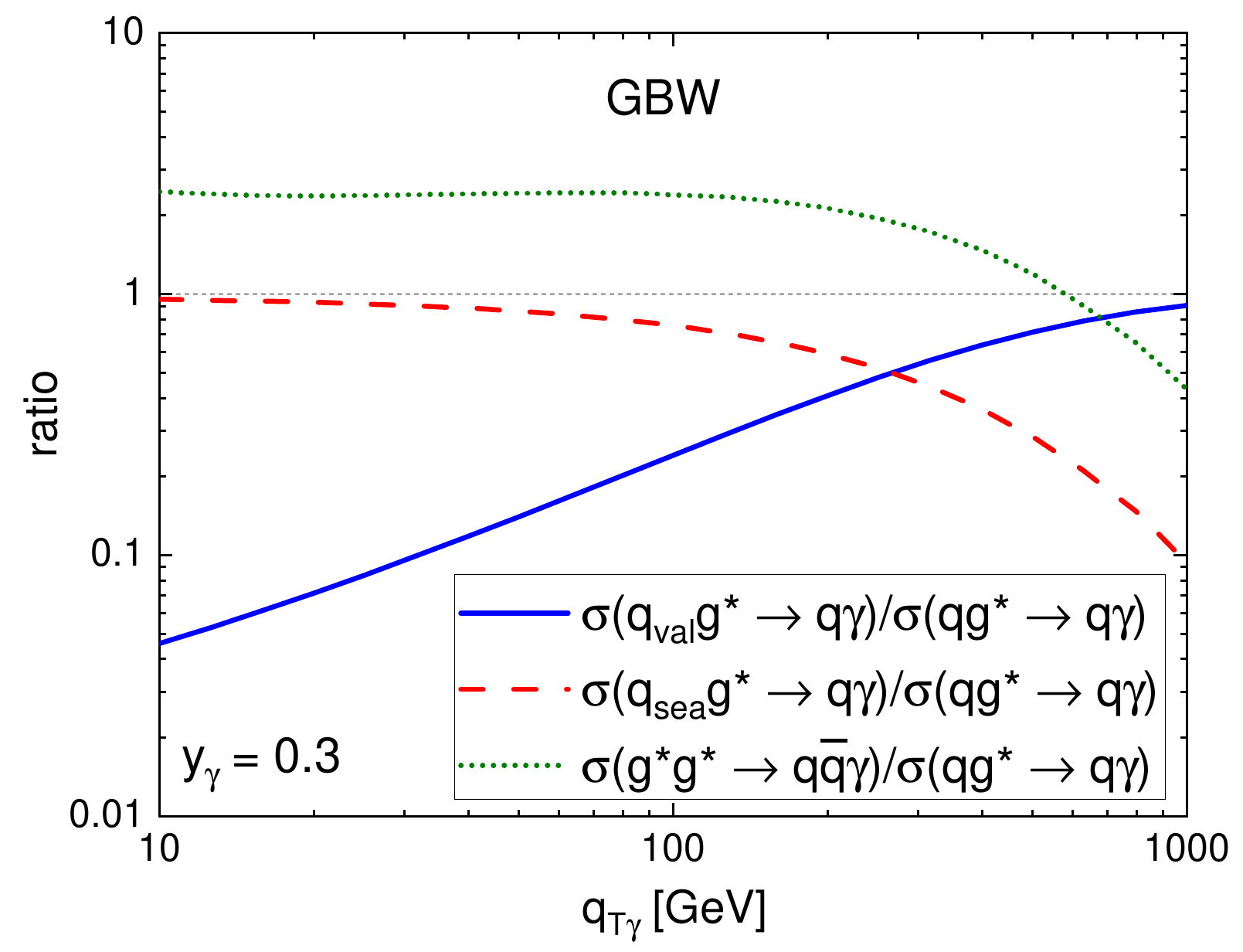}
\hskip -3mm
\includegraphics[width=0.35\textwidth]{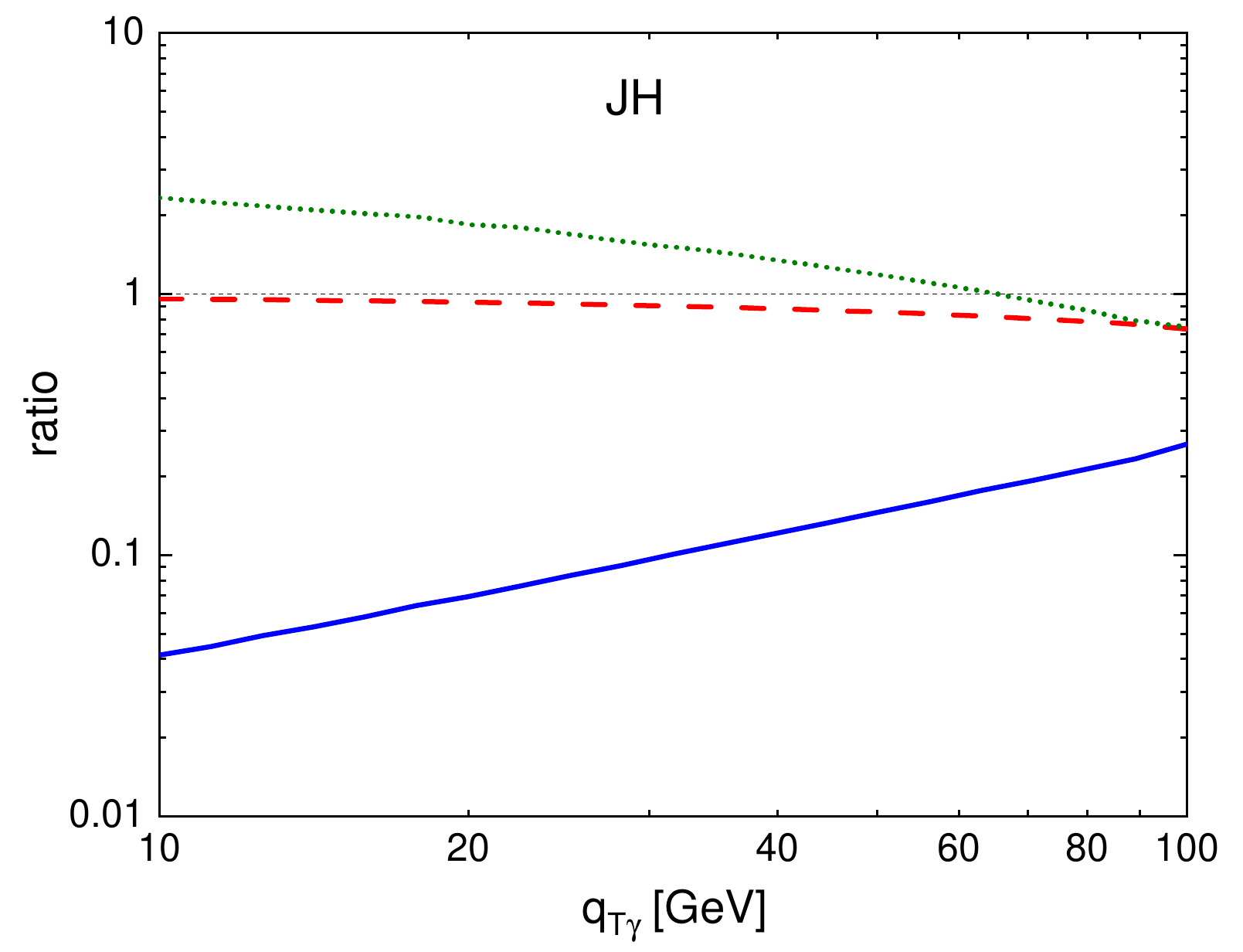}
\hskip -3mm
\includegraphics[width=0.35\textwidth]{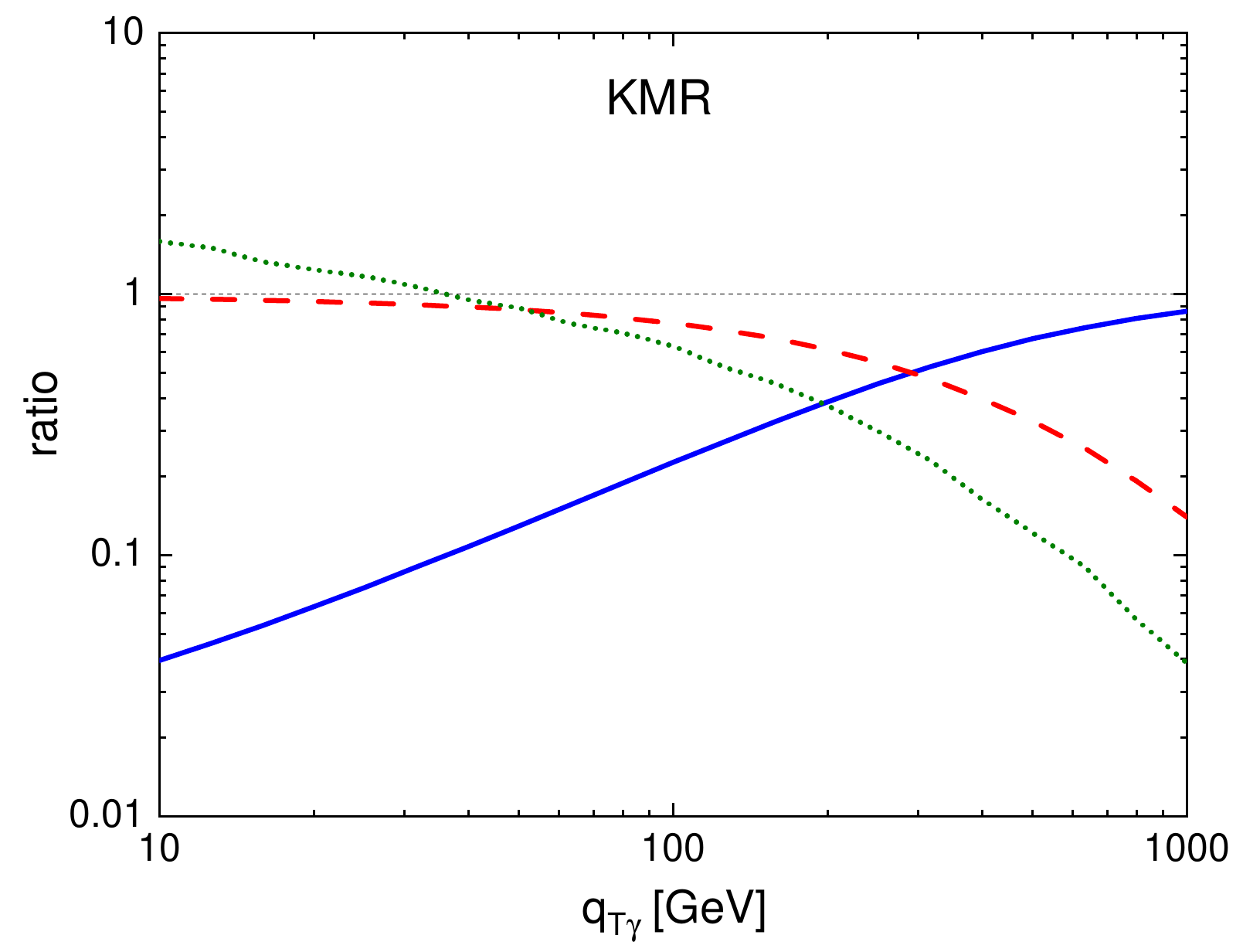}
\end{tabular}
\caption{
Decomposition of the prompt photon production cross section into partonic channels in $qg^*$ and $g^*g^*$ schemes.
Plotted ratios of the partonic cross sections: $\sigma(q_{\mathrm{val}}g^*) / \sigma(qg^*)$,  $\sigma(q_{\mathrm{sea}}g^*) / \sigma(qg^*)$, and $\sigma(g^*g^*) / \sigma(qg^*)$.}
\label{fig:partonic_ratios}
\end{figure}

Let us move now to the other approach,  the $g^*g^*$ scheme, which results are shown in Fig.~\ref{fig:comp_ATLAS_CMS_78ratio_gg}.
Here the cross sections are obtained by adding the $g^*g^* \to q\bar q \gamma$ and $q_{\mathrm{val}}\,g^* \to q\gamma$ contributions. Due to different dependencies on $x$ of the valence quark and gluon distributions, the  $g^*g^* \to q\bar q \gamma$ is dominant at lower $q_T$ while $q_{\mathrm{val}}\,g^* \to q\gamma$ takes over at larger $q_T$. In the region where the gluon $x$ is small or moderate, i.e.\ for  $q_T < 50~{\rm GeV}$, the $g^*g^*$ scheme gives a rather good description of all the data sets with the KMR-AO gluon.  The scale uncertainty is significantly larger than it was for the $qg^*$ scheme. This approach, however, is consistently less successful for $q_T>50$~GeV. The biggest deviation occurs in the region of $q_T \sim ~300-500~{\rm GeV}$, where the theory to experiment ratio is at the level of  $0.6-0.7$. For larger $q_T$,  the theory results become dominated by the $q_{\mathrm{val}}\,g^* \to q\gamma$ channel, and the results approach the results of the $qg^*$ scheme, see Fig.~\ref{fig:partonic_ratios}. Within its limits of applicability, the JH gluon gives rather similar results to the KMR-AO gluon, hence well consistent with the data. The approach based on the GBW gluon experiences strengthened problems due to inaccuracy of the used extrapolation beyond $x>0.01$ and too mild scale dependence since the gluon distribution enters the hadronic cross sections twice in the $g^*g^*$ scheme.

There is an interesting message coming from the comparison of the results in the $qg^*$ anf $g^*g^*$ schemes. On the diagrammatic level, the $g^* g^*$ contribution incorporates the $q_{\mathrm{sea}} \,g^*$ contribution, where the sea quark is produced in the last splitting. Therefore, one expects that the $g^*g^*$ scheme should lead to results that are close to the $qg^*$ scheme. This is, however,  not confirmed by the data at larger $q_T$. In order to provide more insight into this problem, in Fig.~\ref{fig:partonic_ratios} we show the partonic components of the cross sections in both the schemes. We plot the ratios: $\sigma(q_{\mathrm{val}}g^*) / \sigma(qg^*)$, $\sigma(q_{\mathrm{sea}}g^*) / \sigma(qg^*)$ and $\sigma(g^*g^*) / \sigma(qg^*)$, where the reference cross section $\sigma(qg^*) = \sigma(q_{\mathrm{sea}}g^*) +  \sigma(q_{\mathrm{val}}g^*)$ is obtained in the $qg^*$ scheme. Note that $\sigma(q_{\mathrm{val}}g^*)$ is the same in both schemes. 

Clearly, at lower $q_T$, that is for smaller values of the parton $x$, the $q_{\mathrm{sea}} \,g^*$ or $g^*g^*$ contributions strongly dominate  over the $q_{\mathrm{val}}\,g^*$ channel. For $q_T \sim 200-600~{\rm GeV}$, depending on the scheme, the $q_{\mathrm{val}}\,g^*$ channel is leading. Interestingly enough, both for the  KMR-AO and  JH gluons, the $g^*g^*$ channel is larger at the lowest $q_T=10~{\rm GeV}$, but it decreases faster with $q_T$ than the 
$q_{\mathrm{sea}}\,g^*$ channel, and at $q_T \sim 1000~{\rm GeV}$ the difference is already very pronounced. At this stage, we do not have good understanding of this behavior. One could attempt to connect this deviation to going out of the small~$x$ domain, but this simple explanation is undermined by the fact that the results obtained in the $qg^*$ scheme stay close to data even at $q_T \sim 1000~$GeV, where the gluon~$x$ is not small. Hence we consider this problem to be an interesting theoretical puzzle that calls for explanation.

\section{Conclusions}
\label{Sec:5}

In this paper, we have analyzed the prompt photon hadroproduction at the LHC using the $k_T$-factorization approach with the $qg^* \to q\gamma$ and $g^*g^* \to q\bar q\gamma$ partonic channels.The data from the ATLAS and CMS collaboration were considered, obtained at  $\sqrt{S}=7$~TeV and $\sqrt{S}=8$~TeV with the range of the photon transverse momentum from $15~{\rm GeV}$ to $1500~{\rm GeV}$ in several rapidity bins of the photon. The unintegrated, transverse momentum dependent gluon distributions of different origin were probed:  from the integral KMR procedure with angular ordering (KMR-AO), the Jung--Hansson distribution (JH) obtained from the CCFM equation as well as  the Jung-Hautmann newer version JH-2013
and the  gluon distribution from the GBW saturation model extended to large values of $x$. 

With the $qg^*$ partonic channel and  the KMR-AO gluon distribution,  the best overall description of the data was obtained with typical accuracy of $10-20\%$. The results obtained with the JH gluon were found to be slightly less accurate in the region of their applicability with $q_T<100~{\rm GeV}$. The JH-2013 results significantly overshoot the data. 
The GBW gluon does not provide satisfactory description beyond this region and for
central values of the photon rapidity, which shows the expected region of its validity due to small-$x$ nature of this distribution.

We point out that our findings indicate that the precision prompt photon data allow for constraining the transverse momentum distribution of the gluons in the proton. At larger photon rapidities,  the JH description of the data deteriorates for increasing photon transverse momentum $q_T$, underestimating the data while the KMR-AO description stays close to the data. This effect may be traced back to a much steeper decrease of the JH distribution than the KMR-AO one for gluon transverse momenta $k_T$ larger than the factorization scale $\mu_F$.

The description obtained with the $g^*g^*$ channel was shown to provide a good description of the data at lower photon transverse momentum $q_T$, where also the values of the partons' $x$ are lower. At larger $q_T$, however,  the data are significantly underestimated in this approach. We find this result rather puzzling as the $g^*g^*$ channel should  partially include the NLO effects besides the sea quark contributions to the $qg^*$ channel from the last splitting, which is the main contribution to the total $qg^*$ channel. This puzzle calls for a complete NLO analysis of the prompt photon hadroproduction in the $k_T$-factorization framework.

\section*{Acknowledgments}
We thank Błażej Ruba for performing an independent calculation of the $g^*g^*$ matrix elements. Support of the the National Science Center, Poland, Grants Nos. 2017/27/B/ST2/02755, \\ 2019/33/B/ST2/02588 and 2019/32/C/ST2/00202 is gratefully acknowledged. This project has received funding from the European Union's Horizon 2020 research and innovation programme under grant agreement No 824093.


\bibliographystyle{h-physrev4}
\bibliography{mybib_1}

\end{document}